\documentclass{amsart}
\usepackage[dvips]{graphicx}
\usepackage{epsfig,enumerate,mathrsfs}

\usepackage{color}

\numberwithin{equation}{section}
\newtheorem{theorem}{Theorem}[section]
\newtheorem{proposition}[theorem]{Proposition}
\newtheorem{lemma}[theorem]{Lemma}

\newtheorem{remark}[theorem]{Remark}

\DeclareMathOperator{\Tr}{Tr}

\newcommand{\la}{\lambda}

\newcommand{\R}{\mathbb R}
\newcommand{\Z}{\mathbb Z}

\def\N{\mathbb{N}}
\def\C{\mathbb{C}}
 
\def\Id{\mathrm{Id}}

\def\KD3{\mathrm{KD}^3}
\def\dD{\mathscr{D}}
\def\max{\mathrm{max}}
\def\Ei{\mathrm{Ei}}
\begin{document}

\title[a very unusual zeta function]{The very
unusual properties of the resolvent, heat kernel, and zeta
function for the operator   $-d^2/dr^2 - 1/(4r^2)$}
%{\tiny (Journal of Mathematical Physics; Final Version)}

\author{Klaus Kirsten}
\address{Department of Mathematics\\ Baylor University\\
         Waco\\ TX 76798\\ U.S.A. }
\email{Klaus$\_$Kirsten@baylor.edu}

\author{Paul Loya}
\address{Department of Mathematics \\Binghamton University\\
Vestal Parkway East\\Binghamton\\NY 13902\\ U.S.A. }
\email{paul@math.binghamton.edu}

\author{Jinsung Park}
\address{School of Mathematics\\ Korea Institute for Advanced Study\\
207-43\\ Cheongnyangni 2-dong\\ Dongdaemun-gu\\ Seoul 130-722\\
Korea } \email{jinsung@kias.re.kr}

\begin{abstract}
In this article we analyze the resolvent, the heat kernel and the
spectral zeta function of the operator $-d^2/dr^2 - 1/(4r^2)$ over
the finite interval. The structural properties of these spectral
functions depend strongly on the chosen self-adjoint realization
of the operator, a choice being made necessary because of the
singular potential present. Only for the Friedrichs realization
standard properties are reproduced, for all other realizations
highly nonstandard properties are observed. In particular, for
$k\in \N$ we find terms like $(\log t)^{-k}$ in the small-$t$
asymptotic expansion of the heat kernel. Furthermore, the zeta
function has $s=0$ as a logarithmic branch point.
\end{abstract}

\maketitle

%%%%%%%%%%%%%%%%%%%%%%%%%%%%%%%%%%%%%%%%%%%%%%%%%%%%%%%%%%%%%%%%%
\section{Introduction}
%%%%%%%%%%%%%%%%%%%%%%%%%%%%%%%%%%%%%%%%%%%%%%%%%%%%%%%%%%%%%%%%%

%***************************************************************%
\subsection{Zeta functions and an unusual example}
%***************************************************************%

It is well-known that the zeta function of a Laplacian over a
smooth compact manifold, with or without boundary, defines a
meromorphic function on $\C$ with simple poles at prescribed
half-integer values depending on the dimension of the manifold
\cite{BGiP95}. (For a manifold with boundary, we put local
boundary conditions, e.g.\ Dirichlet conditions.) These properties
have far reaching applications in physics as well as mathematics,
e.g. in the context of Casimir energies, effective actions and
analytic torsion; see, for example,
\cite{dowk76-13-3224,gerald,BElE-etc94,HawS77,BKirK01,RS71}.

Surprisingly, there is a completely natural example of a zeta
function for which the described properties break down and which
has no meromorphic extension to $\C$. Let $\Omega \subset \R^2$ be
any compact region and take polar coordinates $(x,y)
\longleftrightarrow (r,\theta)$ centered at any fixed point in
$\Omega$. Then in these coordinates, the standard Laplacian on
$\R^2$ takes the form
\[
\Delta_{\R^2} = - \partial_x^2 - \partial_y^2 = - \partial_r^2 -
\frac{1}{r} \partial_r - \frac{1}{r^2} \partial_\theta^2,
\]
and the measure transforms to $d x d y = r d r d \theta$. A short
computation shows that
\[
\Delta_{\R^2} \phi = \left( - \partial_r^2 - \frac{1}{r}
\partial_r - \frac{1}{r^2} \partial_\theta^2 \right) \phi = \mathcal{R}^{-1}
\Big(- \partial_r^2 + \frac{1}{r^{2}} \big( -
\partial_\theta^2 - \frac14 \big) \Big) \mathcal{R}{\phi},
\]
where $\mathcal{R}$ is the multiplication map by $r^{1/2}$, which
is an isometry from $L^2(\Omega, rdr d \theta)$ to  $L^2(\Omega,
dr d \theta)$. Hence, the following two operators are equivalent
under $\mathcal{R}$:
\[
\Delta_{\R^2} \quad  \longleftrightarrow \quad - \partial_r^2 +
\frac{1}{r^{2}}A,\qquad \text{where}\quad A = -
\partial_\theta^2 - \frac14 .
\]
In the zero eigenspace of $-\partial_\theta^2$, we obtain the
operator of the form $- \partial_r^2 - \frac{1}{4 r^{2}}$. Then
this Laplace type operator has many different self-adjoint
realizations parameterized by angles $\theta \in [0,\pi)$; the
angle $\theta = \pi/2$ corresponds to the so-called Friedrichs
realization. Each realization has a discrete spectrum
\cite{BLeM97}. Consider any one of the realizations, say
$\Delta_\theta$, with $\theta \ne \pi/2$ and form the
corresponding zeta function
\[
\zeta(s ,\Delta_\theta) := \sum_{\la_j \ne 0} \frac{1}{\la_j^s},
\]
where the $\la_j$'s are the eigenvalues of $\Delta_\theta$. The
shocking fact is that \emph{every} such zeta function
corresponding to an angle $\theta \in [0,\pi)$, except $\theta =
\pi/2$, does not have a meromorphic extension to $\C$; in fact
each such zeta function has a logarithmic branch cut with $s=0$ as
the branch point.

%***************************************************************%
\subsection{Self-adjoint realizations}
%***************************************************************%

The properties of the Laplace operator {considered above} boil
down to the main object of consideration in this paper,
\begin{equation} \label{mainopr}
 \Delta
:=-\frac{d^2}{dr^2} - \frac{1}{4 r^{2}} \qquad\text{over}\quad
[0,R] .
\end{equation}
In Section \ref{sec-maxdom} we work out an explicit description of
the maximal domain of $\Delta$. In order to choose a self-adjoint
realization of $\Delta$, we first fix a boundary condition for
$\Delta$ at $r = R$; it turns out that any such boundary condition
for $\phi \in \dD_\max(\Delta)$ must be of the form (see Section
\ref{sec-Lagsubsps})
\begin{equation} \label{domain}
 \cos \theta_2 \, \phi'(R) + \sin \theta_2 \, \phi (R) = 0.
\end{equation}
In other words, the boundary conditions we can choose at $r = R$
are parameterized by angles $\theta_2 \in [0,\pi)$. Note that the
Dirichlet condition is when $\theta_2 = \pi/2$ and the Neumann
condition is when $\theta_2 = 0$. Let us henceforth \emph{fix} an
angle $\theta_2 \in [0,\pi)$ and consider $\Delta$ with the
condition in \eqref{domain}. At $r = 0$, the operator $\Delta$ is
singular {and a limiting procedure $r\to 0$ must be used to define
boundary conditions.} As shown in Section \ref{sec-SAE} (see also
Section \ref{sec-Lagsubsps}), the self-adjoint realizations of
$\Delta$ with the condition \eqref{domain} are again parameterized
by angles $\theta_1 \in [0,\pi)$; the article by Kochube{\u\i}
\cite{KocA79} is perhaps one of the earliest references to contain
such a parameterization. It turns out that $\theta_1 = \pi/2$
corresponds to the Friedrichs realization.

As we will show in Theorem \ref{thneed}, $\phi \in
\dD_\max(\Delta)$ if and only if it can be written in the form
\begin{equation}
\phi = c_1(\phi) \, r^{1/2} + c_2(\phi) \, r^{1/2} \log r +
{\mathcal O} (r^{3/2}) \nonumber ,
\end{equation}
where $c_1(\phi)$ and $c_2(\phi)$ are constants depending on
$\phi$. In terms of these constants, given angles
$\theta_1,\theta_2 \in [0,\pi)$, we consider the operator
\[
\Delta_L := \Delta : \dD_L \to L^2([0,R])
\]
where
\[
\dD_L = \{ \phi \in \dD_\max(\Delta)\, |\, \cos \theta_1 \,
c_1(\phi) + \sin \theta_1\, c_2(\phi) = 0\ , \ \cos \theta_2 \,
\phi'(R) + \sin \theta_2\, \phi(R) = 0 \} .
\]
Here, the subscript ``$L$'' represents the two-dimensional
subspace $L \subset \C^4$ defined by
\[
L := \{(z_1, z_2, z_3, z_4) \in \C^4\ |\ \cos \theta_1 \, z_1 +
\sin \theta_1\, z_2 = 0\ , \ \cos \theta_2 \, z_3 + \sin
\theta_2\, z_4 = 0\}.
\]
This vector space is a Lagrangian subspace of $\C^4$ with respect
to a natural Hermitian symplectic form intimately related to
self-adjoint realizations of $\Delta$; see Section \ref{sec-SAE}.
For general references on this relation see
\cite{HarM00,HarMI00,KaB-StU91,KocA79,KocA90,KocA91,KoV-ScR99}.
For a study of adjoints of ``cone operators'' (in the sense of
Schulze \cite{Sz91}) see \cite{GM}. For properties of heat kernels
and resolvents of cone operators see, for example, \cite{G},
\cite{GKM}, \cite{K}, \cite{SS}.

%***************************************************************%
\subsection{The resolvent, heat kernel, and zeta function}
%***************************************************************%

When $\theta_1 = \pi/2$ (the Friedrichs realization), the
following properties concerning the resolvent, heat kernel, and
zeta function are well known; see for example, Br\"uning and
Seeley \cite{BrJ-SeR87}, Falomir \emph{et al.}\
\cite{FaH-MuM-PiP04}, or Mooers \cite{MoE99}. With $\theta_1 =
\pi/2$, the following properties hold:

\begin{theorem}[Cf.\ \cite{BrJ-SeR87,FaH-MuM-PiP04,MoE99}]
\label{thm-known} Fixing a boundary condition \eqref{domain} at $r
= R$, let $\Delta_L$ denote the corresponding Friedrichs
realization (that is, take $\theta_1 =\pi/2$). Then
\begin{enumerate}[(1)]
\item Let $\Lambda \subset \C$ be any sector (solid angle) not
intersecting the positive real axis. Then as $|\la| \to \infty$
with $\la \in \Lambda$, we have
\[
\Tr( \Delta_L - \la)^{-1} \sim \sum_{k = 1}^\infty a_k \, (-
\la)^{-k/2}.
\]
\item As $t \to 0$, we have
\[
\Tr( e^{-t \Delta_L}) \sim \sum_{k = 0}^\infty \beta_k \, t^{(k -
1)/2} .
\]
\item The zeta function
\[
\zeta(s,\Delta_L) = \Tr (\Delta_L^{-s})
\]
extends from $\Re s > 1/2$ to a meromorphic function on $\C$ with
poles at $s = 1/2 - k$ for $k = 0,1,2,\ldots$.
\end{enumerate}
\end{theorem}

These properties are ``usual'' in that they remain valid, with
appropriate changes, to Laplace-type operators on compact
manifolds (with or without boundary); see for example Gilkey's
book \cite{BGiP95} for a thorough treatment. The first result of
this paper shows that for \emph{any} other realization, these
properties are completely destroyed.

\begin{theorem} \label{thm-main}
With any boundary condition \eqref{domain} fixed at $r = R$,
choose a self-adjoint realization $\Delta_L$ of the resulting
operator that is \textbf{not} the Friedrichs realization. (That
is, take $\theta_1 \ne \pi/2$). Let $\kappa = \log 2 - \gamma -
\tan \theta_1$ where $\gamma$ is the Euler constant. Then the
following properties hold:
\begin{enumerate}[(1)]
\item Let $\Lambda \subset \C$ be any sector (solid angle) not
intersecting the positive real axis. Then as $|\la| \to \infty$
with $\la \in \Lambda$, we have {
\[
\Tr( \Delta_L - \la)^{-1} \sim \frac 1 {(-\la ) (\log (-\la) - 2
\kappa)}
%(- \la)^{-1} \sum_{k = 0}^\infty a_k \, (\log (-\la))^{-k- 1}
+ \sum_{k = 1}^\infty b_k \, (- \la)^{-k/2}.
\]}
%where $a_k = (2 \kappa)^{k}$ for $k = 0,1,2,\ldots$  (in
%particular, $a_0 = 1 \ne 0$).

\item As $t \to 0$, we have (here $\Im$
denotes ``imaginary part of'')
\[
\Tr( e^{-t \Delta_L}) \sim  \frac 1 \pi \Im \left(
\int\limits_1^\infty e^{-tx} \frac 1 {x (\log x+i\pi -2\kappa
)}\,\, dx\right) + \sum_{k = 0}^\infty \beta_k \, t^{(k-1)/2} .
\]

%\[
%\Tr( e^{-t \Delta_L}) \sim \sum_{k = 1}^\infty \alpha_k (\log
%t)^{-k} + \sum_{k = 0}^\infty \beta_k \, t^{(k-1)/2} .
%\]
%with the $\alpha_k$'s depending on $\kappa$ via (here $\Im$
%denotes ``imaginary part of'')
%\[
%\alpha_k = - \frac{1}{k \pi} \Im \left( \int_0^\infty e^{- x}
%\Big(\log x + i \pi - 2 \kappa \Big)^k d x \right),\quad k =
%1,2,3,\ldots.
%\]
\item The zeta function $\zeta(s,\Delta_L)$ can be written in the
form
\[
\zeta(s,\Delta_L) = - \frac{e^{- 2 s \kappa} \sin \pi s}{\pi} \log
s + \zeta_L(s),
\]
where $\zeta_L(s)$ extends from $\Re s > 1/2$ to a holomorphic
function on $\C$ with poles at $s = 1/2 - k$ for $k =
0,1,2,\ldots$. In particular, $\zeta(s , \Delta_L)$ has $s = 0$ as
a logarithmic branch point!
\end{enumerate}
\end{theorem}

\begin{remark}
\em The authors have never seen a natural \emph{geometric}
differential operator with discrete spectrum on a compact manifold
having a spectral zeta function with properties of the sort
described in this theorem.
\end{remark}

\begin{remark}
\em The first term in assertion (1) can be expanded further if
needed. In the formulation of this theorem we leave it in this
more useful compact form.
\end{remark}

\begin{remark} \label{rem-exp}
\em The same kind of remark holds for the first term in assertion
(2). Expanding further we obtain the following expansion: As $t
\to 0$, we have
\begin{equation} \label{exp}
\Tr( e^{-t \Delta_L}) \sim \sum_{k = 1}^\infty \alpha_k (\log
t)^{-k} + \sum_{k = 0}^\infty \beta_k \, t^{(k-1)/2}
\end{equation}
with the $\alpha_k$'s depending on $\kappa$ via (here $\Im$
denotes ``imaginary part of'')
\[
\alpha_k = - \frac{1}{k \pi} \Im \left( \int_1^\infty e^{- x}
\Big(\log x + i \pi - 2 \kappa \Big)^k d x \right),\quad k =
1,2,3,\ldots.
\]
The expansion \eqref{exp} is misleading as written because for $k>
1$, the terms $\beta_k t^{(k-1)/2}$ are sub-leading to any of the
inverse log terms. However, we interpret the first sum in the
expansion \eqref{exp} to mean that for all $N$, we have
\[
\frac 1 \pi \Im \left( \int\limits_1^\infty e^{-tx} \frac 1 {x
(\log x+i\pi -2\kappa )}\,\, dx\right) = \sum_{k = 1}^{N} \alpha_k
(\log t)^{-k} + \mathcal{O}\Big( (\log t)^{-N-1} \Big) .
\]
\end{remark}

%\begin{remark}
%The same kind of remark holds for the first term in assertion (2).
%In addition to the behavior already explicitly displayed,
%expanding further we obtain the terms $\sum_{k=1}^\infty \alpha_k
%(\log t)^{-k}$ with the $\alpha_k$'s depending on $\kappa$ via
%\[
%\alpha_k = - \frac{1}{k \pi} \Im \left( \int_0^\infty e^{- x}
%\Big(\log x + i \pi - 2 \kappa \Big)^k d x \right),\quad k =
%1,2,3,\ldots.
%\]
%For $k> 1$, the terms $\beta_k t^{(k-1)/2}$ in assertion (2) are
%sub-leading to any of these terms.
%\end{remark}

%As the referee pointed out, Theorem \ref{thm-main} can be
%obtained using a parametrix of the resolvent of the corresponding
%realization on $(0,\infty)$ to handle the singularity at $0$ (see
%\cite{BrJ-SeR85} for the case of the Friedrichs realization) and
%patching this together with a parametrix near $r = R$. However, we
%follow a different approach developed in
%\cite{BoM-GeB-KiK-ElE96,BKD,BKD1,KM03} using implicit eigenvalue
%equations. The merit of this approach is that we can directly
%obtain properties of $\z$-functions and explicit formulae for
%$\z$-determinants, which is the second main result of this paper
%described as follows.
%***************************************************************%
\subsection{Explicit formula for the zeta determinant}
%***************************************************************%

Our second result is an explicit formula for a regularized
determinant. For concreteness, we shall impose the Dirichlet
boundary condition at $r = R$. That is, given an angle $\theta \in
[0,\pi)$ with $\theta \ne \pi/2$, we consider the operator
\[
\Delta_\theta := \Delta : \dD_\theta \to L^2([0,R])
\]
where
\[
\dD_\theta = \{ \phi \in \dD_\max(\Delta)\, |\, \cos \theta \,
c_1(\phi) + \sin \theta \, c_2(\phi) = 0\ , \phi(R) = 0 \} .
\]
Then from Theorem \ref{thm-main}, the zeta function $\zeta(s ,
\Delta_\theta)$ has the following form
\[
\zeta(s,\Delta_\theta) = - \frac{e^{- 2 s \kappa} \sin \pi s}{\pi}
\log s + \zeta_\theta(s),
\]
where $\zeta_\theta(s)$ extends from $\Re s > 1/2$ to a
holomorphic function on $\C$ with poles at $s = 1/2 - k$ for $k =
0,1,2,\ldots$. In particular, $\zeta(s , \Delta_\theta)$ has the
form
\begin{equation} \label{zetaslogs}
\zeta(s, \Delta_\theta) \sim - s \log s + \mathcal{O}(s^2 \log s)
+ \text{holomorphic} \qquad \text{as $s \to 0$.}
\end{equation}
In particular,
\[
\zeta'(s, \Delta_\theta) \sim - \log s + \mathcal{O}(s \log s) +
\text{holomorphic} \qquad \text{as $s \to 0$,}
\]
so the $\zeta$-regularized determinant $\det (\Delta_\theta) :=
\exp( - \zeta'(0,\Delta_\theta))$ is \emph{not} defined! However,
from \eqref{zetaslogs}, we see that
\[
\zeta_{\mathrm{reg}}(s,\Delta_\theta) := \zeta(s , \Delta_\theta)
+ s \log s
\]
does have a well-defined derivative at $s = 0$. For this reason,
we define
\[
{\det}_{\mathrm{reg}} (\Delta_\theta) := \exp \Big( -
\zeta_{\mathrm{reg}}'(0,\Delta_\theta) \Big) .
\]
In the following theorem, we give a beautiful explicit formula for
this regularized determinant.

\begin{theorem} \label{thm-main2}
For any $\theta \in [0,\pi)$ with $\theta \ne \pi/2$, we have
\[
{\det}_{\mathrm{reg}} (\Delta_\theta) = \begin{cases} 2 \sqrt{2
\pi R} \, e^\gamma (\tan \theta -\log R) &   \tan \theta\neq \log R\\
\sqrt{ \frac{\pi R}{2}} \, e^\gamma R^2 &  \tan \theta=\log R .
\end{cases}
\]
\end{theorem}

We remark that when $\theta = \pi/2$, the zeta function $\zeta(s,
\Delta_\theta)$ is regular at $s = 0$ and we can also compute the
(usual) $\zeta$-regularized determinant: For $\theta = \pi/2$, we
have
\[
\det (\Delta_\theta) = \sqrt{2 \pi R} ,
\]
a well known result, see e.g.\ Theorem 2.3 of \cite{matthias},
Proposition 5.2 of \cite{LMP}.

We now outline this article. In Sections
\ref{sec-maxdom}--\ref{sec-Lagsubsps} we study the self-adjoint
realizations of our main operator using the Hermitian symplectic
theory due to Gelfand \cite[p.\ 1]{INovS99}; cf.\ also
\cite{FaH-MuM-PiP04,KocA79,KocA90,KocA91,LMP,MoE99,PavB87}.
Although some of this material can be found piecemeal throughout
the literature, we present all the details here in order to keep
our article elementary, self-contained, and ``user-friendly". In
Sections \ref{sec-eigen}--\ref{sec-heat} we prove Theorem
\ref{thm-main} in the special case that the Dirichlet boundary
condition is chosen at $r = R$ and in Section \ref{sec-det} we
prove Theorem \ref{thm-main2}, all using the contour integration
method developed in \cite{BoM-GeB-KiK-ElE96,BKD,BKD1}. In Section
\ref{sec-bdycond} we prove Theorem \ref{thm-main} in full
generality. Finally, in Appendix \ref{sec-appendix}, we explicitly
calculate the resolvent of $\Delta_\theta$, which is needed at
various places in our analysis.
%Finally, in Appendix \ref{sec-appendix} we show how
%the resolvent method of Falomir \emph{et al.} \cite{FaH-MuM-PiP04}
%can be used to derive Theorem \ref{thm-main}.

The authors express their sincere gratitude to the referee for his
or her kind suggestions, exhortations, and careful readings, which
greatly helped us to improve the exposition and quality of this
paper.

%%%%%%%%%%%%%%%%%%%%%%%%%%%%%%%%%%%%%%%%%%%%%%%%%%%%%%%%%%%%%%%%%
\section{The maximal domain} \label{sec-maxdom}
%%%%%%%%%%%%%%%%%%%%%%%%%%%%%%%%%%%%%%%%%%%%%%%%%%%%%%%%%%%%%%%%%

Our first order of business is to characterize the self-adjoint
realizations of the operator in \eqref{mainopr}; for general
references on self-adjoint realizations and their applications to
physics see, e.g.,
\cite{AuJ-JaU-SkV95,BoJ-PuJ03,BuW-GeF85,falo03-36-9991,FaH-PiP01,FrW-LaD-SpR71,
KocA79,KocA90,KocA91,KoV-ScR99,PavB87,ChT-FuT-TsI03}. To do so, we
first need to determine the maximal domain of $\Delta$:
\[
\dD_\max(\Delta) := \{ \phi \in L^2([0,R])\, |\, \Delta \phi \in
L^2([0,R])\} .
\]
For a quick review, $\Delta \phi$ is understood in the
distributional sense; thus, $\Delta \phi$ is the functional on
test functions $C_c^\infty ((0,R))$ defined by
\[
(\Delta \phi ) (\xi) := \int_0^R \Delta \xi (r) \, \overline{\phi
(r)}\, dr \quad \text{for all} \quad \xi \in C_c^\infty ((0,R)).
\]
Then $\Delta \phi \in L^2([0,R])$ means that the distribution
$\Delta \phi : C_c^\infty ((0,R)) \to \C$ is represented by an
$L^2$ function in the sense that there is a function $f \in
L^2([0,R])$ such that
\[
\int_0^R \Delta \xi (r) \, \overline{\phi (r)}\, dr = \langle \xi
, f \rangle \quad \text{for all} \quad \xi \in C_c^\infty ((0,R))
\]
where $\langle \cdot , \cdot \rangle$ denotes the $L^2$ inner
product (conjugate linear in the second slot) on $L^2([0,R])$. The
following theorem is inspired by Falomir \emph{et al.} \cite[Lem.\
2.1]{FaH-MuM-PiP04}.

\begin{theorem} \label{thneed} $\phi \in \dD_\max(\Delta)$ if and only if $\phi$ can be
written in the form
\begin{equation} \label{phic12}
\phi = c_1(\phi) \, r^{1/2} + c_2(\phi) \, r^{1/2} \log r +
\widetilde{\phi},
\end{equation}
where $c_1(\phi) , c_2(\phi)$ are constants and $\widetilde{\phi}$
is a continuously differentiable function on $[0,R]$ such that
$\widetilde{\phi}(r)= \mathcal{O}(r^{3/2})$, $\widetilde{\phi}'(r)
= \mathcal{O}(r^{1/2})$, and $\Delta \widetilde{\phi} \in
L^2([0,R])$.
\end{theorem}
\begin{proof} Since
\[
\Delta \big( c_1\, r^{1/2} + c_2\, r^{1/2} \log r \big) = 0,
\]
it follows that any $\phi$ of the stated form is in
$\dD_\max(\Delta)$. Now let $\phi \in \dD_\max(\Delta)$; then
$\Delta \phi = f \in L^2([0,R])$. Let us define $\psi := r^{-1/2}
\phi$ so that $\phi = r^{1/2} \psi$. Then
\[
f = - \phi'' - \frac{1}{4 r^2} \phi = \frac14 r^{-3/2} \psi -
r^{-1/2} \psi' - r^{1/2} \psi'' - \frac14 r^{-3/2} \psi = -
r^{-1/2} \psi' - r^{1/2} \psi'' .
\]
After multiplication by $r^{1/2}$, we get
\[
\psi' + r \psi'' = - r^{1/2} f \quad \Longrightarrow \quad (r
\psi')' = - r^{1/2} f.
\]
Since $r^{1/2}$ and $f$ are in $L^2([0,R])$, by the
Cauchy-Schwartz inequality, we know that $r^{1/2} f$ is in
$L^1([0,R])$, therefore we can conclude that
\begin{equation} \label{psi'}
\psi' = \frac{c_2}{r} - \frac{1}{r} \int_0^r t^{1/2}\, f(t)\, dt .
\end{equation}
Notice that by Cauchy-Schwartz,
\begin{equation} \label{estimatet12f}
\left| \int_0^r t^{1/2}\, f(t)\, dt \right| \leq \sqrt{\int_0^r
t\, dt} \cdot \|f\|_2 = \frac{r}{\sqrt{2}} \|f\|_2.
\end{equation}
Thus, the second term on the right in \eqref{psi'} is in
$L^1([0,R])$. Therefore, from \eqref{psi'} we see that
\[
\psi(r) = c_1 + c_2\, \log r - \int_0^r \frac{1}{x} \int_0^x
t^{1/2}\, f(t)\, dt\, dx,
\]
or, since $\phi = r^{1/2} \psi$, we get
\[
\phi(r) = c_1\, r^{1/2} + c_2\, r^{1/2} \log r + \widetilde{\phi}\
\ , \qquad \widetilde{\phi} : = - r^{1/2} \int_0^r \frac{1}{x}
\int_0^x t^{1/2}\, f(t)\, dt\, dx .
\]
By \eqref{estimatet12f}, we have
\[
\left| \int_0^r \frac{1}{x} \int_0^x t^{1/2}\, f(t)\, dt\, dx
\right| \leq \int_0^r \frac{1}{\sqrt{2}} \|f\|_2\, dx =
\frac{r}{\sqrt{2}} \|f\|_2.
\]
From this estimate, it follows that $\widetilde{\phi}(r) =
\mathcal{O}(r^{3/2})$ and $\widetilde{\phi}'(r) =
\mathcal{O}(r^{1/2})$.
\end{proof}

%%%%%%%%%%%%%%%%%%%%%%%%%%%%%%%%%%%%%%%%%%%%%%%%%%%%%%%%%%%%%%%%%
\section{Self-adjoint realizations} \label{sec-SAE}
%%%%%%%%%%%%%%%%%%%%%%%%%%%%%%%%%%%%%%%%%%%%%%%%%%%%%%%%%%%%%%%%%

Choosing a linear subspace $\dD \subset \dD_\max(\Delta)$, we say
that
\[
\Delta_\dD := \Delta : \dD \to L^2([0,R])
\]
is \emph{self-adjoint} (in which case $\Delta_\dD$ is called a
\emph{self-adjoint realization} of $\Delta$) if
\[
\{ \psi \in \dD_\max(\Delta)\ |\ \langle \Delta \phi , \psi
\rangle = \langle \phi , \Delta \psi \rangle \ \ \text{for all}\ \
\phi \in \dD \} = \dD;
\]
in other words, $\Delta$ is symmetric on $\dD$ and adding any
elements to $\dD$ will destroy this symmetry.

In order to determine if $\Delta$ has any self-adjoint
realization, we need to analyze the quadratic form
\[
\langle \phi , \Delta \psi \rangle - \langle \Delta \phi , \psi
\rangle\quad \text{for}\ \ \phi,\psi \in \dD_\max(\Delta).
\]
It turns out that this difference is related to finite-dimensional
symplectic linear algebra. Let us define
\[
\omega : \C^4 \times \C^4 \to \C
\]
by \begin{equation}\label{symp}
\omega(v,w) : = v_1\,
\overline{w_2} - v_2 \, \overline{w_1} + v_3\, \overline{w_4} -
v_4\, \overline{w_3}.
\end{equation}
The function $\omega$ is Hermitian antisymmetric and
non-degenerate; for this reason, $\omega$ is called a
\emph{Hermitian symplectic form}.

\begin{theorem} \label{thm-quadform}
Let $\phi,\psi \in \dD_\max(\Delta)$ be written in the form
\eqref{phic12}, i.e.\
\[
\phi = c_1(\phi) \, r^{1/2} + c_2(\phi) \, r^{1/2} \log r +
\widetilde{\phi},
\]
where $\widetilde{\phi}$ is continuously differentiable with
$\widetilde{\phi}(r)= \mathcal{O}(r^{3/2})$, $\widetilde{\phi}'(r)
= \mathcal{O}(r^{1/2})$, and $\Delta \widetilde{\phi} \in
L^2([0,R])$, and with a similar formula holding for $\psi$. Then,
\[
\langle \phi , \Delta \psi \rangle - \langle \Delta \phi , \psi
\rangle = \omega(\vec{\phi} ,\vec{\psi}),
\]
where $\omega$ is the Hermitian symplectic form defined above and
$\vec{\phi}, \vec{\psi} \in \C^4$ are the vectors
\[
\vec{\phi} := (c_1(\phi), c_2(\phi) , \phi'(R), \phi (R)) \ \ , \
\ \vec{\psi} := (c_1(\psi), c_2(\psi) , \psi'(R), \psi (R)) .
\]
\end{theorem}
\begin{proof}
We have
\begin{align}
\notag \langle \phi , \Delta \psi \rangle - & \langle \Delta \phi
, \psi \rangle  = \lim_{\varepsilon \to 0} \int_\varepsilon ^R
\Big( \phi(r)\, \overline{\Delta \psi (r)}\, - \Delta
\phi(r) \, \overline{\psi(r)} \Big) \, dr \\
%\notag & = \lim_{\varepsilon \to 0} \int_\varepsilon ^R \Big( -
%\phi(r)\, \overline{\psi'' (r)}\, +
%\phi''(r) \, \overline{\psi(r)} \Big) \, dr \\
\notag & = \lim_{\varepsilon \to 0} \int_\varepsilon ^R
\frac{d}{dr} \Big( - \phi(r)\, \overline{\psi' (r)}\, +
\phi' (r) \, \overline{\psi(r)} \Big) \, dr \\
\label{limsae} & = \lim_{\varepsilon \to 0} \Big(
\phi(\varepsilon)\, \overline{\psi' (\varepsilon)}\, - \phi'
(\varepsilon) \, \overline{\psi(\varepsilon)} \Big) + \Big( \phi'
(R) \, \overline{\psi(R)} - \phi(R)\, \overline{\psi' (R)} \Big) .
\end{align}
Recall that
\[
\phi = c_1(\phi) \, r^{1/2} + c_2(\phi) \, r^{1/2} \log r +
\widetilde{\phi}\ \ , \ \ \psi = c_1(\psi) \, r^{1/2} + c_2(\psi)
\, r^{1/2} \log r + \widetilde{\psi},
\]
where $\widetilde{\phi}$ and $\widetilde{\psi}$ are continuously
differentiable functions on $[0,R]$ such that $\widetilde{\phi}(r)
,\widetilde{\psi} (r)$ $=$ $\mathcal{O}(r^{3/2})$,
$\widetilde{\phi}'(r) ,\widetilde{\psi}'(r) =
\mathcal{O}(r^{1/2})$. Taking derivatives, we get
\[
\phi' = \frac{c_1(\phi)}{2} \, r^{-1/2} + \frac{c_2(\phi)}{2}
r^{-1/2} (\log r + 2) + \widetilde{\phi} '
\]
and similarly for $\psi '$.
%\[
%\psi' = \frac{c_1(\psi)}{2} \, r^{-1/2} + \frac{c_2(\psi)}{2}
%r^{-1/2} (\log r + 2) + \widetilde{\psi} '.
%\]
It follows that
\begin{multline*}
\phi(\varepsilon)\, \overline{\psi' (\varepsilon)} =
\frac{c_1(\phi)\, \overline{c_1(\psi)}}{2} + \frac{c_1(\phi) \,
\overline{c_2(\psi)}}{2} (\log \varepsilon + 2) \\ +
\frac{c_2(\phi)\, \overline{c_1(\psi)}}{2} \, \log \varepsilon +
\frac{c_2(\phi) \, \overline{c_2(\psi)}}{2} \log \varepsilon (\log
\varepsilon + 2) + o(1)
\end{multline*}
and similarly for $\phi ' (\varepsilon)\,
\overline{\psi(\varepsilon)}$.
%\begin{multline*}
%\phi' (\varepsilon) \, \overline{\psi(\varepsilon)} =
%\frac{c_1(\phi)\, \overline{c_1(\psi)}}{2} + \frac{c_1(\phi) \,
%\overline{c_2(\psi)}}{2} \log \varepsilon \\ + \frac{c_2(\phi)\,
%\overline{c_1(\psi)}}{2} \, (\log \varepsilon + 2) +
%\frac{c_2(\phi) \, \overline{c_2(\psi)}}{2} \log \varepsilon (\log
%\varepsilon + 2) + o(1).
%\end{multline*}
Subtracting, we get
\[
\phi(\varepsilon)\, \overline{\psi' (\varepsilon)}\, - \phi'
(\varepsilon) \, \overline{\psi(\varepsilon)} = c_1(\phi)\,
\overline{c_2(\psi)} - c_2(\phi)\, \overline{c_1(\psi)} + o(1).
\]
Combining this with \eqref{limsae} proves our result.
\end{proof}

Recall that a subspace $L \subset \C^4$ is called
\emph{Lagrangian} if $L^{\perp_\omega} = L$ where
$L^{\perp_\omega}$ is the orthogonal complement of $L$ with
respect to $\omega$; explicitly, $L$ is Lagrangian means
\[
\{ w \in \C^4\ |\ \omega( v , w) = 0 \ \ \text{for all}\ \ v \in L
\} = L.
\]
We now have our main result.

\begin{theorem} \label{thm-sae}
Self-adjoint realizations of $\Delta$ are in one-to-one
correspondence with Lagrangian subspaces of $\C^4$ in the sense
that given any Lagrangian subspace $L \subset \C^4$, defining
\[
\dD_L := \{ \phi \in \dD_\max(\Delta) \ |\ \vec{\phi} \in L\}
\]
the operator
\[
\Delta_L := \Delta : \dD_L \to L^2([0,R])
\]
is self-adjoint and any self-adjoint realization of $\Delta$ is of
the form $\dD_L$ for some Lagrangian $L \subset \C^4$.
\end{theorem}
\begin{proof}
By definition,
\[
\Delta_\dD := \Delta : \dD \to L^2([0,R])
\]
is \emph{self-adjoint} means
\[
\{ \psi \in \dD_\max(\Delta)\ |\ \langle \Delta \phi , \psi
\rangle = \langle \phi , \Delta \psi \rangle \ \ \text{for all}\ \
\phi \in \dD \} = \dD.
\]
By Theorem \ref{thm-quadform}, we can write this as: $\Delta_\dD$
is self-adjoint if and only if
\begin{equation} \label{Lagcondi}
\omega(\vec{\phi}, \vec{\psi}) = 0 \ \ \text{for all}\ \ \phi \in
\dD \quad \Longleftrightarrow \quad \psi \in \dD.
\end{equation}

Suppose that $\Delta_\dD$ is self-adjoint and define $L := \{
\vec{\phi} \in \C^4\ |\ \phi \in \dD\}$; we shall prove that $L$
is Lagrangian. Let $w \in L$ and choose $\psi \in \dD$ such that
$\vec{\psi} = w$. Then by \eqref{Lagcondi}, $\omega(\vec{\phi} ,
w) = 0$ for all $\phi \in \dD$. Therefore, $\omega(v,w) = 0$ for
all $v \in L$. Conversely, let $w \in \C^4$ and assume that
$\omega(v,w) = 0$ for all $v \in L$. Choose $\psi \in
\dD_\max(\Delta)$ such that $\vec{\psi} = w$; e.g.\ if $w =
(w_1,w_2,w_3,w_4)$, then
\begin{equation} \label{psiconstruct}
\psi := w_1\, r^{1/2} + w_2\, r^{1/2} \log r + (w_3 - w_4) (r - R)
+ w_4 r
\end{equation}
will do. Then $\omega(v,w) = 0$ for all $v \in L$ implies that
$\omega(\vec{\phi} , \vec{\psi}) = 0$ for all $\phi \in \dD$,
which by \eqref{Lagcondi}, implies that $\psi \in \dD$, which
further implies that $w = \vec{\phi} \in L$.

Now let $L \subset \C^4$ be Lagrangian; we shall prove that
$\Delta_L$ is self-adjoint, that is, \eqref{Lagcondi} holds. Let
$\psi \in \dD_L$. Then, since $L$ is Lagrangian, we automatically
have $\omega(\vec{\phi} , \vec{\psi}) = 0$ for all $\phi \in
\dD_L$. Conversely, let $\psi \in \dD_\max(\Delta)$ and assume
that $\omega(\vec{\phi} , \vec{\psi}) = 0$ for all $\phi \in
\dD_L$. By the construction \eqref{psiconstruct} given any $v \in
L$ we can find a $\phi \in \dD_\max(\Delta)$ such that $\vec{\phi}
= v$. Therefore, $\omega(\vec{\phi} , \vec{\psi}) = 0$ for all
$\phi \in \dD_L$ implies that $\omega(v , \vec{\psi}) = 0$ for all
$v \in L$, which by the Lagrangian condition on $L$, implies that
$\vec{\psi} \in L$. This shows that $\psi \in \dD_L$ and our proof
is complete.
\end{proof}

%%%%%%%%%%%%%%%%%%%%%%%%%%%%%%%%%%%%%%%%%%%%%%%%%%%%%%%%%%%%%%%%%
\section{More on Lagrangian subspaces} \label{sec-Lagsubsps}
%%%%%%%%%%%%%%%%%%%%%%%%%%%%%%%%%%%%%%%%%%%%%%%%%%%%%%%%%%%%%%%%%

The symplectic form $\omega : \C^4 \times \C^4 \to \C$ defined in
\eqref{symp} is naturally separated into two parts:
\begin{equation} \label{omega00}
\omega(v,w) = \omega_0((v_1,v_2),(w_1,w_2)) +
\omega_0((v_3,v_4),(w_3,w_4))
\end{equation}
where
\[
\omega_0 : \C^2 \times \C^2 \to \C\ \ \text{is defined by}\ \
\omega_0(v,w) = v_1\, \overline{w_2} - v_2 \, \overline{w_1}.
\]
The first $\omega_0$ appearing in \eqref{omega00} corresponds to
the singularity at $r=0$ and the second $\omega_0$ in
\eqref{omega00} corresponds to the boundary $r = R$. For this
reason, it is natural to focus on Lagrangian subspaces $L \subset
\C^4$ of the form $L = L_1 \oplus L_2$ where $L_i \subset \C^2$ is
Lagrangian with respect to $\omega_0$. With this in mind, let us
characterize all such Lagrangian subspaces of $\C^2$. First, we
observe that

\begin{lemma} We can write
\[
\omega_0(v,w) = \langle G v , w \rangle\quad \text{for all}\ \ v,w
\in \C^2,
\]
where $\langle \ , \ \rangle$ denotes the inner product on $\C^2$
and $G = \begin{pmatrix} 0 & - 1 \\ 1 & 0 \end{pmatrix}$.

\end{lemma}

Recalling that $L \subset \C^2$ is Lagrangian means that
\[
\{ w \in \C^2\ |\ \omega_0( v , w) = 0 \ \ \text{for all}\ \ v \in
L \} = L,
\]
from this lemma, it is straightforward to show that
\[
L \subset \C^2 \ \ \text{is Lagrangian if and only if}\ \ G
L^\perp = L,
\]
where $L^\perp$ is the orthogonal complement of $L$ with respect
to the inner product $\langle \ , \ \rangle$.  From this, one can
easily prove the following main result in this section.

\begin{theorem} \label{thm-Ltheta}
$L \subset \C^2$ is Lagrangian if and only if $L = L_\theta$ for
some $\theta \in \R$ where
\[
L_\theta = \{ (x,y) \in \C^2\ |\ \cos \theta\, x + \sin \theta\, y
= 0 \}.
\]
\end{theorem}

Notice that we can restrict to $0 \leq \theta < \pi$ in Theorem
\ref{thm-Ltheta}. Let $\theta_1, \theta_2$ be two such angles and
put $L := L_{\theta_1} \oplus L_{\theta_2}$. As in \eqref{phic12},
we write $\phi \in \dD_\max(\Delta)$ as
\[
\phi = c_1(\phi) \, r^{1/2} + c_2(\phi) \, r^{1/2} \log r +
\widetilde{\phi},
\]
where $\widetilde{\phi}$ is continuously differentiable with
$\widetilde{\phi}(r)= \mathcal{O}(r^{3/2})$, $\widetilde{\phi}'(r)
= \mathcal{O}(r^{1/2})$, and $\Delta \widetilde{\phi} \in
L^2([0,R])$. Then as a consequence of Theorem \ref{thm-sae}, we
know that
\begin{equation} \label{DeltaL}
\Delta_L := \Delta : \dD_L \to L^2([0,R])
\end{equation}
is self-adjoint, where
\[
\dD_L = \{ \phi \in \dD_\max(\Delta)\, |\, \cos \theta_1 \,
c_1(\phi) + \sin \theta_1\, c_2(\phi) = 0\ , \ \cos \theta_2 \,
\phi'(R) + \sin \theta_2\, \phi(R) = 0 \} .
\]
When $\theta_1 = \pi/2$, then we are requiring $c_2(\phi)$ vanish
so that near $r = 0$, we have
\[
\phi = c_1(\phi) \, r^{1/2} + \widetilde{\phi};
\]
that is, no log terms; in \cite{BrJ-SeR87}, Br\"uning and Seeley
prove that $\theta_1 = \pi/2$ is the Friedrichs realization of the
operator $\Delta$ acting on smooth functions supported away from
$r = 0$ with the boundary condition $\cos \theta_2 \, \phi'(R) +
\sin \theta_2\, \phi(R) = 0$ at $r = R$. As seen in Theorem
\ref{thm-known} in the Introduction, this self-adjoint realization
gives rise to the ``usual'' resolvent, heat kernel, and zeta
function properties. When $\theta_1 \ne \pi/2$, we get very
pathological properties as shown in Theorem \ref{thm-main}. In the
following sections we enter in the proof of Theorem
\ref{thm-main}.

%%%%%%%%%%%%%%%%%%%%%%%%%%%%%%%%%%%%%%%%%%%%%%%%%%%%%%%%%%%%%%%%%
\section{Eigenvalues with Dirichlet conditions at $r = R$}
\label{sec-eigen}
%%%%%%%%%%%%%%%%%%%%%%%%%%%%%%%%%%%%%%%%%%%%%%%%%%%%%%%%%%%%%%%%%

As shown in detail in Section \ref{sec-bdycond}, the strange
behaviors depicted in Theorem \ref{thm-main} do not depend on the
choice of the Lagrangian $L_2$ (that is, the choice of boundary
condition at $r = R$). For this reason, we shall use $\theta_2 =
0$ for the Lagrangian $L_2$ in \eqref{DeltaL}; thus, we shall
consider the self-adjoint operator $\Delta_\theta := \Delta :
\dD_\theta \to L^2([0,R])$, where $0 \leq \theta < \pi$ and
$\theta \ne \pi/2$, and
\begin{align*}
\dD_\theta  = \{ \phi \in \dD_\max(\Delta)\, |\, \cos \theta \,
c_1(\phi) + \sin \theta\, c_2(\phi) = 0\ , \ \phi(R) = 0 \};
\end{align*}
so we are simply imposing the Dirichlet condition at $r = R$.

We now find {an explicit formula for the eigenfunctions and} a
transcendental equation, which determines the spectrum of
$\Delta_\theta$. We begin with the following eigenvalue equation:
\[
(\Delta_\theta - \mu^2) \phi = 0 \quad \Longleftrightarrow \quad
\phi'' + \frac{1}{4 r^{2}}\phi + \mu^2 \phi = 0.
\]
We can turn this into a Bessel equation via the usual trick by
setting $\phi = r^{1/2} \psi(\mu r)$. Then,
\[
\phi'' = - \frac14 r^{-3/2} \psi(\mu r) + \mu r^{-1/2} \psi'(\mu
r) + \mu^2 r^{1/2} \psi''(\mu r),
\]
so
\[
\phi'' + \frac{1}{4 r^{2}}\phi + \mu^2 \phi = 0 \quad
\Longleftrightarrow \quad \mu r^{-1/2} \psi'(\mu r) + \mu^2
r^{1/2} \psi''(\mu r) + \mu^2 r^{1/2} \psi(\mu r) = 0,
\]
or
\[
(\mu r)^2\, \psi''(\mu r) + (\mu r)\, \psi'(\mu r) + (\mu r)^2\,
\psi(\mu r) = 0 .
\]
For fixed $\mu$, the solutions to this equation are linear
combinations of $J_0$ and $Y_0$ (with $Y_0$ the Bessel function of
the second kind), so
\[
\phi = C_1 r^{1/2} J_0(\mu r) + C_2 r^{1/2} Y_0(\mu r).
\]
Using that \cite[p.\ 360]{BAbM-StI92}
\begin{equation} \label{Y0def}
\frac{\pi}{2} Y_0(z) := \big( \log z - \log 2 + \gamma \big)
J_0(z) - \sum_{k = 1}^\infty \frac{H_k (- \frac14 z^2 )^k}{(k!)^2}
,
\end{equation}
where $H_k := 1 + \frac12 + \cdots + \frac{1}{k}$, the form
(\ref{phic12}) for $\phi\in\dD_\max(\Delta)$ is obtained by
choosing the constants $C_1$ and $C_2$ in such a way that
\[
\phi = c_1(\phi)\, r^{1/2} \, J_0(\mu r) +c_2(\phi)\, r^{1/2} \,
\Big(\frac{\pi}{2}  Y_0(\mu r) - (\log \mu - \log 2 + \gamma)\,
J_0(\mu r)\Big) .
\]
By definition of the Bessel function \cite[p.\ 360]{BAbM-StI92},
we have as $z \to 0$,
\begin{align}\label{Jvdef}
J_v(z) & = \frac{z^v}{2^v} \sum_{k = 0}^\infty \frac{(-\frac14
z^2)^k}{k! \, \Gamma(v + k + 1)} \\ & = \frac{z^v}{2^v \Gamma(1 +
v)} \hspace{-.2em} \left( 1 - \frac{z^2}{4 (1 + v)} +
\frac{z^4}{32 (1 + v) (2 + v)} - + \cdots \right)\notag
\end{align}
and by \eqref{Y0def}, we see that
\[
\phi = c_1(\phi)\, r^{1/2} + c_2(\phi)\, r^{1/2} \, \log r +
\mathcal{O}((\mu r)^2),
\]
where $\mathcal{O}((\mu r)^2)$ is a power series in $(\mu r)^2$
vanishing like $(\mu r)^2$ as $r \to 0$. Therefore, by definition
of $\dD_\theta$, we have
\begin{equation} \label{bdycond0}
\cos \theta \, c_1(\phi) + \sin \theta \, c_2 (\phi) = 0.
\end{equation}
To satisfy the Dirichlet condition at $r = R$, we must have
\[
c_1(\phi)\, J_0(\mu R) +c_2(\phi)\, \Big(\frac{\pi}{2} Y_0(\mu R)
- (\log \mu - \log 2 + \gamma)\, J_0(\mu R)\Big) = 0 .
\]
It follows that
\[
\det \begin{pmatrix} \cos \theta & \sin \theta \\ J_0(\mu R) &
\frac{\pi}{2} Y_0(\mu R) - (\log \mu - \log 2 + \gamma)\, J_0(\mu
R )
\end{pmatrix} = 0,
\]
or $\frac{\pi}{2} Y_0(\mu R) - (\log \mu - \log 2 + \gamma)\,
J_0(\mu R) = \tan \theta\, J_0(\mu R)$. We summarize our findings
in the following proposition.

\begin{proposition} The transcendental equation
\begin{equation} \label{eigeneqn}
F(\mu) := \frac{\pi}{2} Y_0(\mu R) - (\log \mu - \kappa)\, J_0(\mu
R ) = 0 \ , \qquad \kappa = \log 2 - \gamma - \tan \theta
\end{equation}
determines the eigenvalues of $\Delta_\theta$.
\end{proposition}
In the following theorem we state various properties of the
eigenvalues of $\Delta_\theta$; note that in \cite[p.\
4572]{FaH-MuM-PiP04} it is stated that there are no negative
eigenvalues; however, it turns out that for example when $\pi/2 <
\theta < \pi$ and $R\geq 1$, there is \emph{always} a negative
eigenvalue.

\begin{theorem} \label{thm-eigenv}For $0 \leq \theta < \pi$ with $\theta \ne \frac \pi 2 $,
\begin{enumerate}[(1)]
\item $\Delta_\theta$ has a zero eigenvalue if and only if $\ \log
R= \tan \theta$. \item $\Delta_\theta$ has a unique negative
eigenvalue if and only if $\tan \theta < \log R$.
%if and only if
%$R\geq 1$ and $\frac\pi 2 < \theta < \pi$ or if $\ 0<R<1$ and
%$\frac\pi 2 < \theta < \pi + \arctan \log R$.
\end{enumerate}
\end{theorem}
\begin{proof} Using \eqref{Y0def} and the expansion
\begin{equation}\label{e:J0}
J_0(z) = \sum_{k = 0}^\infty \frac{(-\frac14 z^2)^k}{ (k!)^2},
\end{equation}
the eigenvalue equation $\frac{\pi}{2} Y_0(\mu R) - (\log \mu -
\log 2 + \gamma)\, J_0(\mu R) = \tan \theta\, J_0(\mu R)$ can be
written as
\begin{equation} \label{eveq}
- \sum_{k = { 1 }}^\infty \frac{H_k (- \frac14 \mu^2 R^2
)^k}{(k!)^2} = (-\log R+ \tan \theta) \sum_{k = 0}^\infty
\frac{(-\frac14 \mu^2R^2)^k}{ (k!)^2}.
\end{equation}
Thus, $\mu = 0$ solves this equation if and only if $\log R = \tan
\theta$.

Now $\Delta_\theta$ has a negative eigenvalue means that $\mu = i
x$ for $x$ real solves \eqref{eveq}:
\begin{equation} \label{eveq2}
- \sum_{k = 1}^\infty \frac{H_k (\frac14 x^2R^2 )^k}{(k!)^2}
=(-\log R +\tan \theta) \sum_{k = 0}^\infty \frac{( \frac14 x^2
R^2)^k}{ (k!)^2}.
\end{equation}
If $(-\log R +\tan \theta) > 0$, then \eqref{eveq2} has no
solutions because the right-hand side of \eqref{eveq2} will be
strictly positive for all real $x$ while the left-hand side of
\eqref{eveq2} is nonpositive. Thus, we may assume that $\alpha
:=\log R - \tan \theta > 0$.
%which restricts $R$ and $\theta$ to
%the values given in assertion {\it (2)}.
Then we can write \eqref{eveq2} as
\[
f(x R) = 0 \quad \text{where}\ \ f(x) = \sum_{k = 1}^\infty
\frac{H_k\, x^{2 k}}{4^k(k!)^2} - \sum_{k = 0}^\infty \frac{\alpha
x^{2 k}}{4^k (k!)^2};
\]
thus, we just have to prove that $f(x) = 0$ has a unique solution.
To prove this, observe that since the harmonic series $1 + \frac12
+ \frac13 + \cdots$ diverges, we can choose $N \in \N$ such that
$H_N > \alpha > H_{N-1}$. We now write
\[
f(x) = \sum_{k = N}^\infty \frac{(H_k - \alpha) \, x^{2 k}}{4^k
(k!)^2} - \left( \sum_{k = 0}^{N-1} \frac{(\alpha - H_k)\, x^{2
k}}{4^k (k!)^2} \right),
\]
where $H_0 := 0$, and note that $f(x) = 0$ if and only if $g(x) =
0$ where
\[
g(x) := x^{-2 N} f(x) = \sum_{k = N}^\infty \frac{(H_k - \alpha)
\, x^{2 (k-N)}}{4^k (k!)^2} - \left( \sum_{k = 0}^{N-1}
\frac{(\alpha - H_k)}{4^k (k!)^2 \, x^{2(N- k)}} \right) .
\]
Because of the powers of $x$ in the denominator the second sum on
the right, we see that $g(0+) = - \infty$ while because of the
first sum on the right, we see that $\lim_{x \to \infty} g(x) =
\infty$. In particular, by the intermediate value theorem, $g(x) =
0$ for some $0 < x < \infty$. Since
\[
g'(x) = \sum_{k = N+1}^\infty \frac{(H_k - \alpha) \, 2(k - N)
x^{2 (k-N)-1}}{4^k (k!)^2} + \left( \sum_{k = 0}^{N-1} \frac{2(N -
k)(\alpha - H_k)}{4^k (k!)^2 \, x^{2(N- k)+1}} \right) > 0
\]
the function $g$ is strictly increasing, so there is only one $x >
0$ such that $g(x) = 0$. It follows that $f(x) = 0$ for a unique
$x > 0$ and our proof is now complete. A graph of $f(x)$ for $R=1$
and $\tan\theta=-2$ is shown in Figure \ref{fig-eigengraph}.
\begin{figure}
\epsfig{file=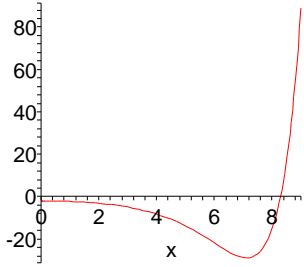, width=1.7in} \caption{Graph of
$f(x)$ when $R=1$ and $\tan\theta=-2$.} \label{fig-eigengraph}
\end{figure}
\end{proof}

%%%%%%%%%%%%%%%%%%%%%%%%%%%%%%%%%%%%%%%%%%%%%%%%%%%%%%%%%%%%%%%%%
\section{The $\zeta$-function with
Dirichlet conditions at $r = R$} \label{sec-zeta}
%%%%%%%%%%%%%%%%%%%%%%%%%%%%%%%%%%%%%%%%%%%%%%%%%%%%%%%%%%%%%%%%%

Let $0 \leq \theta < \pi$ with $\theta \ne \pi/2$. We now analyze
the zeta function using the contour integral techniques developed
in \cite{BKD,BKD1,BKirK01}.

\begin{figure}[b]
\centering \includegraphics{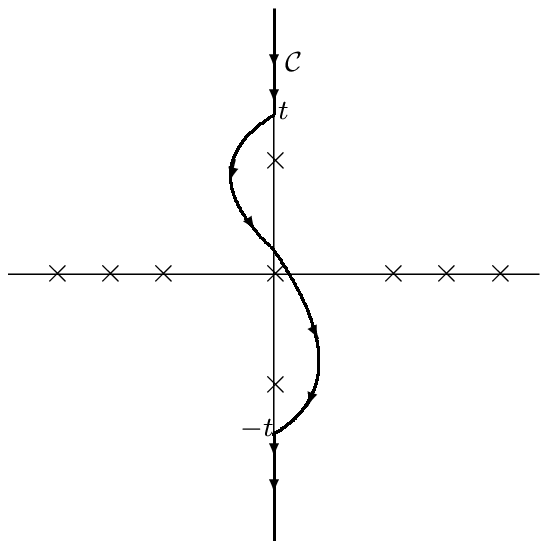} \caption{The
contour $\gamma$ for the zeta function. The $\times$'s represent
the zeros of $F(\mu)$. The squares of the $\times$'s on the
imaginary axis represent the negative eigenvalues of
$\Delta_\theta$. Here, $t$ is on the imaginary axis and is larger
in absolute value than the absolute value of the negative
eigenvalue of $\Delta_\theta$ (if one exists). The contour
$\gamma_t$ goes from $t$ to $-t$.} \label{fig-contstrange1}
\end{figure}

In Appendix \ref{sec-appendix}, Theorem \ref{thm-traceres}, we
have shown that
\begin{align*}
\Tr ( \Delta_\theta - \mu^2)^{-1} & =  - \frac{1}{2 \mu}
\frac{d}{d \mu} \log F(\mu).
\end{align*}
Therefore, for $\Re s>1/2$, by {\it definition} the zeta function
is given by
%By the argument principle the zeta
%function is given by
\begin{equation}\label{zeta-def}
\zeta(s , \Delta_\theta ) = \frac{1}{2 \pi i} \int_\gamma \mu^{-2
s} \frac{d}{d \mu} \log F(\mu) d \mu = \frac{1}{2 \pi i}
\int_\gamma \mu^{-2 s} \frac{F'(\mu)}{F(\mu)} d \mu ,
\end{equation}
where $\gamma$ is a contour in the plane shown in Figure
\ref{fig-contstrange1}.
%{The contour $\gamma$ is obtained as a result of a sequence of
%finite closed curves constructed as follows. Let $R_j>0$, $j\in
%\N$, be a monotonically increasing sequence of positive numbers
%with $\lim_{j\to\infty} R_j = \infty$ and with $R_j \neq \mu_i^2$
%for all $i,j\in\N$. Let $\gamma_j$ be the finite contour similar
%to the one in Figure \ref{fig-contstrange1} but running from
%$iR_j$ to $-iR_j$ only, and with an added semicircle of radius
%$R_j$ in the right half-plane. For increasing $j$ an increasing
%number of eigenvalues is enclosed by $\gamma_j$. By the following
%proposition, as $j\to\infty$, for $\Re s > 1/2$ the contribution
%from the semicircle vanishes and the above integral converges to
%the zeta function $\zeta(s,\Delta_\theta)$.}

To analyze properties of the zeta function, we need the following
technical lemma.
\begin{lemma} \label{lem-asympF}
Let $0 \leq \theta < \pi$ with $\theta \ne \frac\pi 2$ and
$\Upsilon \subset \C$ be a sector (closed angle) in the right-half
plane. Then as $|x| \to \infty$ with $x \in \Upsilon$, we have
\begin{equation} \label{asymF0}
F(i x)\, \sim \, - \frac{1}{\sqrt{2\pi}} (\log x  - \kappa)
(xR)^{-\frac12} e^{xR} \Big( 1 + \frac{1}{8 xR}+ \frac{9}{2(8x
R)^2}+ \mathcal{O}(x^{-3}) \Big) ,
\end{equation}
where $\mathcal{O}(x^{-3})$ is a power series in $x^{-1}$ starting
from $x^{-3}$, and
\begin{equation} \label{asymF}
\frac{d}{d x} \log F(i x) \sim  \frac{1}{x(\log x  - \kappa)} + R
- \frac{1}{2 x} - \frac{1}{8 x^2 R} + \mathcal{O}(x^{-3}) ,
\end{equation}
with the same meaning for $\mathcal{O}(x^{-3})$. Finally, $F(\mu)$
is an even function of $\mu$, and as $\mu \to 0$, we have
\begin{equation} \label{Fasymp00}
F(\mu) \sim  (\log R- \tan \theta )+ \frac14 \mu^2 R^2 ( 1 + \tan
\theta - \log R) + \mathcal{O}(\mu^4) .
\end{equation}
\end{lemma}
\begin{proof}
From \eqref{Y0def}, we have
\begin{align*}
\frac{\pi}{2} Y_0(i x) &
 = \big( \log (i x) - \log 2 + \gamma
\big) J_0(i x) - \sum_{k = 1}^\infty \frac{H_k (- \frac14 (i x)^2
)^k}{(k!)^2}\\
& = \big( \log x + i \frac{\pi}{2} - \log 2 + \gamma \big)
I_0(x) - \sum_{k = 1}^\infty \frac{H_k (\frac14 x^2 )^k}{(k!)^2}\\
& = i \frac{\pi}{2} I_0(x) - K_0(x)  ,
\end{align*}
where $I_0(x)$ is the modified Bessel function of the first kind,
and
\[
K_0(x) := - \big( \log x - \log 2 + \gamma \big) I_0(x) + \sum_{k
= 1}^\infty \frac{H_k (\frac14 x^2 )^k}{(k!)^2}
\]
is the modified Bessel function of the second kind. Therefore,
\begin{align*}
F(i x) & = \frac{\pi}{2} Y_0(i xR) - (\log (i x) - \kappa) J_0(i xR)\\
& = i \frac{\pi}{2} I_0(xR) - K_0(xR) - \Big(\log
 x + i \frac{\pi}{2} - \kappa \Big) I_0(xR)\\
& = - (\log x  - \kappa) I_0(xR) - K_0(xR) .
\end{align*}
By \cite[p.\ 377]{BAbM-StI92}, as $|x| \to \infty$ for $x \in
\Upsilon$, we have
\[
I_0(x) \, \sim \, \frac{e^x}{\sqrt{2\pi x}}\Big(1 + \frac{1}{8 x}+
\frac{9}{2(8x)^2}+ \mathcal{O}(x^{-3})\Big)
\]
where $\mathcal{O}(x^{-3})$ is a power series in $x^{-1}$ starting
with $x^{-3}$; furthermore \cite[p.\ 378]{BAbM-StI92}, as $|x| \to
\infty$ for $x \in \Upsilon$,
\[
K_0(x) \, \sim \, \sqrt{\frac{2}{\pi x}} e^{ - x } \Big(1 -
\frac{1}{8 x} + \frac{9}{2(8x)^2}+ \mathcal{O}(x^{-3})\Big).
\]
Therefore, as $|x| \to \infty$ for $x \in \Upsilon$, we have
\begin{align*}
F(i x) & \, \sim  \, - (\log x  - \kappa) I_0(xR)\\
%& \, \sim \, - (\log x  - \kappa) \frac{e^{xR}}{\sqrt{2\pi
%xR}}\Big(1
%+ \frac{1}{8 xR}+ \frac{9}{2(8xR)^2}+ \mathcal{O}(x^{-3})\Big)\\
& \, \sim \,  - \frac{1}{\sqrt{2\pi}} (\log x  - \kappa)
(xR)^{-\frac12} e^{xR} \Big(1 + \frac{1}{8 xR}+
\frac{9}{2(8xR)^2}+ \mathcal{O}(x^{-3})\Big),
\end{align*}
which proves \eqref{asymF0}. Taking logarithms, we see that as
$|x| \to \infty$ for $x \in \Upsilon$, we have
\[
\log F(i x) \sim c + \log (\log x  - \kappa) - \frac12 \log x + x
R+ \log  \Big(1 + \frac{1}{8 xR}+ \frac{9}{2(8xR)^2}+
\mathcal{O}(x^{-3})\Big),
\]
where $c$ is a constant. Since $\log (1 + z) = z - \frac{z^2}{2} +
\frac{z^3}{3} - + \cdots$, we have
\[
\log F(i x) \sim c + \log (\log x  - \kappa) - \frac12 \log x + xR
+ \frac{1}{8 xR} + \mathcal{O}(x^{-2}).
\]
Taking the derivative of this we get \eqref{asymF}.

To determine the asymptotics as $\mu \to 0$, recalling that
$\kappa = \log 2 - \gamma - \tan \theta$, we see that
\begin{align*}
F(\mu) & = \frac{\pi}{2} Y_0(\mu R) - (\log \mu - \kappa)
J_0(\mu R) \notag \\
& = \frac{\pi}{2} Y_0(\mu R) - (\log \mu - \log 2 + \gamma)
J_0(\mu R) - \tan \theta J_0(\mu R) \notag \\
& = \frac14 \mu^2 R^2 + (\log R- \tan \theta) + (\tan \theta -
\log R) \cdot \frac14 \mu^2 R^2 +
\mathcal{O}(\mu^4) \label{Feven}  \\
& =  (\log R- \tan \theta )+ \frac14 \mu^2 R^2( 1 + \tan \theta -
\log R ) + \mathcal{O}(\mu^4),\notag
\end{align*}
where we used \eqref{Y0def} and \eqref{e:J0} in passing from the
second to the third line. In particular, the second line with
\eqref{Y0def} and \eqref{e:J0} shows that $F(\mu)$ is an even
function of $\mu$. This completes our proof.
\end{proof}

We need one more lemma.

\begin{lemma} \label{lem-logint} We have
\[
\int_{|t|}^\infty x^{-2 s} \frac{1}{x(\log x - \kappa)}\, d x = -
e^{- 2 s \kappa} \log s - e^{- 2 s \kappa} \Big( \gamma + \log (2
(\log |t| - \kappa)) + \mathcal{O}(s)\Big),
\]
where $\mathcal{O}(s)$ is an entire function of $s$ that is
$\mathcal{O}(s)$ at $s = 0$.
\end{lemma}
\begin{proof}
In the integral we assume that $\log |t|
> \kappa$ so that the integral is well-defined.
Now to analyze this integral we make the change of variables $u =
\log x - \kappa$ or $x = e^\kappa\, e^u$, and obtain
\[
\int_{|t|}^\infty x^{-2 s} \frac{1}{x(\log x - \kappa)}\, d x =
e^{- 2 s \kappa} \int_{\log |t| - \kappa}^\infty e^{-2 s u}
\frac{d u}{u} .
\]
Making the change of variables $y = 2 s u$, we get
\[
\int_{|t|}^\infty x^{-2 s} \frac{1}{x(\log x - \kappa)}\, d x =
e^{- 2 s \kappa} \int_{2 s (\log |t| - \kappa)}^\infty e^{- y}
\frac{d y}{y} .
\]
Recall that the \emph{exponential integral} is defined by (see
\cite[Sec.\ 8.2]{BGrI-RyI00})
\[
\Ei(z) := - \int_{-z}^\infty e^{-y} \frac{d y}{y} .
\]
Therefore,
\[
\int_{|t|}^\infty x^{-2 s} \frac{1}{x(\log x - \kappa)}\, d x = -
e^{- 2 s \kappa} \Ei \big(-2 s (\log |t| - \kappa)\big).
\]
Also from \cite[p.\ 877]{BGrI-RyI00}, we have
\[
\Ei(z) = \gamma + \log (- z) + \sum_{k = 1}^\infty \frac{z^k}{k
\cdot k!},
\]
therefore
\begin{align}
\notag \int_{|t|}^\infty x^{-2 s} & \frac{1}{x(\log x - \kappa)}\,
d x = - e^{- 2 s \kappa} \Big( \gamma + \log
(2 s (\log |t| - \kappa)) + \mathcal{O}(s) \Big)\\
\label{intxlogx} & = - e^{- 2 s \kappa} \log s - e^{- 2 s \kappa}
\Big( \gamma + \log (2 (\log |t| - \kappa)) + \mathcal{O}(s)\Big),
\end{align}
where $\mathcal{O}(s)$ is an entire function of $s$ that is
$\mathcal{O}(s)$ at $s = 0$. This completes our proof.
\end{proof}

We now determine the structure of the zeta function.

\begin{proposition} \label{prop-zeta}
The zeta function $\zeta(s,\Delta_\theta)$ can be written in the
form
\[
\zeta(s,\Delta_\theta) = - \frac{e^{- 2 s \kappa} \sin \pi s}{\pi}
\log s + \zeta_\theta(s),
\]
where $\kappa = \log 2 - \gamma - \tan \theta$ and
$\zeta_\theta(s)$ extends from $\Re s > 1/2$ to a holomorphic
function on $\C$ with poles at $s = 1/2 - k$ for $k =
0,1,2,\ldots$. In particular, $\zeta(s , \Delta_\theta)$ has $s =
0$ as a logarithmic branch point!
\end{proposition}
\begin{proof}
Recalling \eqref{zeta-def}, we write
\[
\int_\gamma = - \int_{t}^{0 + i \infty} + \int_{-t}^{0-i \infty} +
\int_{\gamma_t},
\]
where $\gamma_t$ is the part of $\gamma$ from $t$ to $-t$, and
using that
\[
i^{-2s} = ( e^{i \pi/2} )^{-2 s} = e^{- i \pi s}\quad \text{and}
\quad (-i)^{-2s} = ( e^{-i \pi/2} )^{-2 s} = e^{i \pi s},
\]
we obtain the integral
\begin{align*}
\zeta(s,\Delta_\theta ) & = \frac{1}{2 \pi i} \int_\gamma \mu^{-2
s}
\frac{d}{d\mu} \log F(\mu) \, d \mu\\
& = \frac{1}{2 \pi i} \bigg\{ - \int_{|t|}^\infty (i x)^{-2 s}
\frac{d}{d x} \log F(i x)\, d x + \int_{|t|}^\infty (-i x)^{-2 s}
\frac{d}{d x} \log F(-i x)\, d x \bigg\} \\
& \hspace{7.5cm} + \frac{1}{2 \pi i} \int_{\gamma_t} \mu^{-2 s}
\frac{F'(\mu)}{F(\mu)}\, d \mu\\ & = \frac{1}{2 \pi i} \Big( -
e^{- i \pi s} + e^{i \pi s}\Big) \int_{|t|}^\infty x^{-2 s}
\frac{d}{d x} \log F(i x)\, d x + \frac{1}{2 \pi i}
\int_{\gamma_t} \mu^{-2 s} \frac{F'(\mu)}{F(\mu)}\, d \mu ,
\end{align*}
or,
\begin{equation} \label{zetafn}
\zeta(s,\Delta_\theta ) = \frac{\sin \pi s}{\pi} \int_{|t|}^\infty
x^{-2 s} \frac{d}{d x} \log F(i x)\, d x + \frac{1}{2 \pi i}
\int_{\gamma_t} \mu^{-2 s} \frac{F'(\mu)}{F(\mu)}\, d \mu ,
\end{equation}
a formula that will be analyzed in a moment. The second integral
here is over a finite contour so an entire function of $s \in \C$,
so we are left to analyze the analytic properties of the first
integral. To do so, recall from \eqref{asymF} that for $x \to
\infty$, we have
\[
\frac{d}{d x} \log F(i x) \sim \frac{1}{x(\log x  - \kappa)} +
\sum_{k = 0}^\infty \beta_k x^{-k},
\]
for some constants $\beta_k$. Since
\[
\frac{\sin \pi s}{\pi} \int_{|t|}^\infty x^{-2 s - k}\, d x =
\frac{\sin \pi s}{\pi} \frac{x^{-2 s - k + 1}}{-2 s - k + 1}
\bigg|_{x = |t|}^\infty = \frac{\sin \pi s}{\pi} \frac{|t|^{-2 s -
k + 1}}{2 s + k - 1}
\]
which has poles at $s = (1 - k)/2$ for $s \notin \Z$, it follows
that the expansion $\sum_{k = 0}^\infty \beta_k x^{-k}$ will
contribute to the function $\zeta_\theta(s)$ in the statement of
this proposition where $\zeta_\theta(s)$ extends from $\Re s
> 1/2$ to a holomorphic function on $\C$ with poles at $s = 1/2 -
k$ for $k = 0,1,2,\ldots$. Lemma \ref{lem-logint} applied to the
integral
\[
\frac{\sin \pi s}{\pi} \int_{|t|}^\infty x^{-2 s} \frac{1}{x(\log
x - \kappa)}\, d x
\]
now completes our proof.
\end{proof}
{\begin{remark} The existence of the logarithmic branch point at
$s=0$ has been missed in \cite{FaH-MuM-PiP04}. The error occurs in
equation (A13) where certain antiderivatives (specifically
$xY_1(x)$ and $x^2 Y_1^2$) were accidentally set equal to zero at
$s=0$.
\end{remark} }
%%%%%%%%%%%%%%%%%%%%%%%%%%%%%%%%%%%%%%%%%%%%%%%%%%%%%%%%%%%%%%%%%
\section{Trace of the resolvent with Dirichlet conditions at $r = R$}
\label{sec-res}

Using the Theorem \ref{thm-traceres}, we can now prove Theorem
\ref{thm-main} (1) for $\Delta_\theta$. We have chosen to present
the results in a form where the first term has been expanded
further; Theorem \ref{thm-main} (1) is contained in equation
\eqref{asymFix} of the proof and the explanation of the meaning of
the expansion is similar to that found in Remark \ref{rem-exp}.

\begin{proposition} \label{prop-res}
Let $\theta \ne \pi/2$ and $\kappa = \log 2 - \gamma - \tan
\theta$, furthermore let $\Lambda \subset \C$ be any sector (solid
angle) not intersecting the positive real axis. Then as $|\la| \to
\infty$ with $\la \in \Lambda$, we have
\[
\Tr( \Delta_\theta - \la)^{-1} \sim (- \la)^{-1} \sum_{k =
0}^\infty a_k \, (\log (-\la))^{-k - 1} + \sum_{k = 1}^\infty b_k
\, (- \la)^{-k/2},
\]
where $a_k = (2 \kappa)^{k}$ for $k = 0,1,2,\ldots$ (in
particular, $a_0 = 1 \ne 0$).
\end{proposition}
\begin{proof}
Setting $\la = - x^2$ with $x \in \Upsilon \subset \C$ a sector in
the right-half plane, it suffices to prove that as $|x| \to
\infty$ with $x \in \Upsilon$, we have
\[
\Tr( \Delta_\theta + x^2)^{-1} \sim x^{-2} \sum_{k = 0}^\infty
\frac{\kappa^{k}}{2} \, (\log x )^{-k - 1} + x^{-1} \sum_{k =
0}^\infty b_k \, x^{-k},
\]
or after multiplication by $2x$, we just have to prove that
\[
2 x \Tr( \Delta_\theta + x^2)^{-1} \sim x^{-1} \sum_{k = 1}^\infty
\kappa^{k} \, (\log x )^{-k - 1} + \sum_{k = 0}^\infty \beta_k \,
x^{-k},
\]
To prove this, we recall from Theorem \ref{thm-traceres} that
\[
2 x \Tr( \Delta_\theta + x^2)^{-1} = \frac{d}{d x} \log F(i x) .
\]
By Lemma \ref{lem-asympF} (see \eqref{asymF}) we know that as $|x|
\to \infty$ for $x \in \Upsilon$,
\begin{equation} \label{asymFix}
\frac{d}{d x} \log F(i x) \sim \frac{1}{x(\log x  - \kappa)} +
\sum_{k = 0}^\infty \beta_k x^{-k}.
\end{equation}
Finally, the expansion
\begin{multline*}
\frac{1}{x(\log x  - \kappa)} = \frac{1}{x \log x} \cdot
\frac{1}{(1 - \kappa (\log x)^{-1})} \\ = \frac{1}{x \log x}
\sum_{k = 0}^\infty \kappa^k (\log x)^{-k} = x^{-1} \sum_{k =
0}^\infty \kappa^k (\log x)^{-k - 1},
\end{multline*}
concludes our result.
\end{proof}

As shown in the proof, for $|x| \to \infty$ with $x \in \Upsilon$,
where $\Upsilon$ is a sector in the right-half plane, we have
\begin{equation}
\label{2xtrace} 2 x \Tr( \Delta_\theta + x^2)^{-1} = \frac{d}{d x}
\log F(i x) \sim \frac{1}{x(\log x  - \kappa)} + \sum_{k =
0}^\infty \beta_k x^{-k},
\end{equation}
or with $\la = - x^2$, as $|\la| \to \infty$ with $\la \in \Lambda
\subset \C$, a sector not intersecting the positive real axis, we
have
\begin{equation}
\label{ltrace} \Tr( \Delta_\theta - \la)^{-1} \sim
\frac{1}{(-\la)(\log (-\la) - 2 \kappa)} + \sum_{k = 1}^\infty b_k
(-\la)^{-k/2}.
\end{equation}
This fact will be used in the next section.

%%%%%%%%%%%%%%%%%%%%%%%%%%%%%%%%%%%%%%%%%%%%%%%%%%%%%%%%%%%%%%%%%
\section{The heat trace with
Dirichlet conditions at $r = R$} \label{sec-heat}
%%%%%%%%%%%%%%%%%%%%%%%%%%%%%%%%%%%%%%%%%%%%%%%%%%%%%%%%%%%%%%%%%

\begin{figure}
\centering \includegraphics{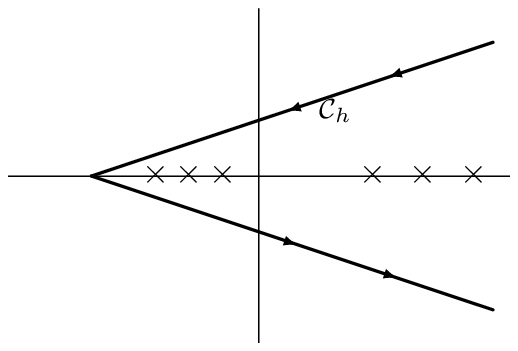} \caption{The
contour $\gamma$.} \label{fig-contstrange}
\end{figure}

To determine the small-time heat asymptotics, we write
\[
\Tr (e^{-t \Delta_\theta} ) = \frac{i}{2 \pi} \int_\gamma e^{-t
\la} \Tr (\Delta_\theta - \la)^{-1} d \la
\]
where $\gamma$ is a counter-clockwise contour in the plane
surrounding the eigenvalues of $\Delta_\theta$; see Figure
\ref{fig-contstrange}. This is the starting point to show Theorem
\ref{thm-main} (2) for $\Delta_\theta$. Again we have chosen to
present the results in a form where the first term has been
expanded further. This makes the actual structure of the small-$t$
expansion more explicit; Theorem \ref{thm-main} (2) is contained
in equation \eqref{int2} of the proof.

\begin{proposition} \label{prop-hk}
As $t \to 0$, we have
\[
\Tr( e^{-t \Delta_\theta}) \sim \sum_{k = 1}^\infty \alpha_k (\log
t)^{-k} + \sum_{k = 0}^\infty \beta_k \, t^{(k-1)/2} .
\]
with the $\alpha_k$'s depending on $\kappa$ via
\[
\alpha_k = - \frac{1}{k \pi} \Im \left( \int_0^\infty e^{- x}
\Big(\log x + i \pi - 2 \kappa \Big)^k d x \right).
\]
\end{proposition}
\begin{proof}
The small-time asymptotics are determined by the large-spectral
parameter asymptotics of the resolvent. Now recall from
\eqref{ltrace} that as $|\la| \to \infty$ with $\la$ in a sector
not intersecting the positive real axis, we have
\[
\Tr( \Delta_\theta - \la)^{-1} \sim \frac{1}{(-\la)(\log (-\la) -
2 \kappa)} + \sum_{k = 1}^\infty b_k (-\la)^{-k/2} .
\]
Since (making the change of variables $\la \mapsto t^{-1}\la$)
\[
\int e^{-t \la} (- \la)^{-k/2} \, d \la = t^{-1 + k/2} \int
e^{-\la} (- \la)^{-k/2} \, d \la
\]
the series $\sum_{k = 1}^\infty b_k \, (- \la)^{-k/2}$ gives rise
to a small time expansion
\[
\sum_{k = 1}^\infty \beta_k \, t^{-1 + k/2} .
\]
Therefore, we just have to analyze the behavior of
\[
\frac{i}{2 \pi} \int_\gamma e^{-t \la} \frac{1}{(-\la)(\log (-\la)
- 2 \kappa)} d \la .
\]
Deforming $\gamma$ to the real line, the integral here is, modulo
a term that is a smooth function of $t$ at $t = 0$, equal to
\[
-\int_1^\infty e^{-t x} \frac{1}{(-x)(\log (-(x + i 0)) - 2
\kappa)} d x + \int_1^\infty e^{-t x} \frac{1}{(-x)(\log (-(x - i
0)) - 2 \kappa)} d x,
\]
or after simplification, this sum becomes
\[
\int_1^\infty e^{-t x} \frac{1}{x(\log x - i \pi - 2 \kappa)} d x
- \int_1^\infty e^{-t x} \frac{1}{x(\log x + i \pi - 2 \kappa)} d
x .
\]
(The reason we start at $x = 1$ is that $1/(x [\log x \pm i\pi - 2
\kappa] )$ is not integrable near $x = 0$.) Since for any complex
number $z$, we have $i (\overline{z} - z) = 2\, \Im z$, we see
that modulo a term that is a smooth function of $t$ at $t = 0$,
\begin{align} \label{int}
\frac{i}{2 \pi} \int_\gamma e^{-t \la} \frac{1}{(-\la)(\log (-\la)
- 2 \kappa)} d \la & \equiv \frac{1}{\pi} \Im \left( \int_1^\infty
e^{-t x} \frac{1}{x(\log x + i \pi - 2 \kappa)} d x \right)\\
\notag & = \frac{1}{\pi} \Im \, \ell(t),
\end{align}
where
\[
\ell(t) := \int_1^\infty e^{-t x} \frac{1}{x(\log x + i \pi - 2
\kappa)} d x .
\]
In summary, we have proved that
\begin{equation} \label{int2}
\Tr( e^{-t \Delta_\theta}) \sim  \frac{1}{\pi} \Im \, \ell(t) +
\sum_{k = 0}^\infty \beta_k \, t^{(k-1)/2} ,
\end{equation}
which is exactly the statement of Theorem \ref{thm-main} (2).

We shall compute the asymptotics of $\ell(t)$ as $t \to 0$. To do
so, observe that
\[
\ell'(t) := - \int_1^\infty e^{-t x} \frac{1}{(\log x + i \pi - 2
\kappa)} d x
\]
Now $1/\log x$ is integrable near $x = 0$, so we can write
\[
\ell'(t) = f(t) + g(t),
\]
where
\[
f(t) := - \int_0^\infty e^{-t x} \frac{1}{(\log x + i \pi - 2
\kappa)} d x \ \ , \ \ g(t) := \int_0^1 e^{-t x} \frac{1}{(\log x
+ i \pi - 2 \kappa)} d x .
\]
Note that $g(t)$ is smooth at $t = 0$. We will now determine the
asymptotics of $f(t)$ near $t = 0$. To this end, we make the
change of variables $x \mapsto t^{-1} x$:
\begin{align*}
f(t) & = - t^{-1} \int_0^\infty e^{- x} \frac{1}{(\log (x/t) + i
\pi - 2 \kappa)} d x \\
%& = - t^{-1} \int_0^\infty e^{- x}
%\frac{1}{(\log x - \log t + i \pi - 2 \kappa)} d x\\
& = t^{-1} (\log t)^{-1} \int_0^\infty e^{- x} \frac{1}{1 -
\frac{\log x + i \pi - 2 \kappa}{\log t} } d x.
\end{align*}
Since $(1 - r)^{-1} = \sum_{k = 0}^N r^k + r^{N+1} (1 - r)^{-1}$
for any $N \in \N$, we see that for any $N \in \N$,
\begin{multline*}
f(t) = t^{-1} (\log t)^{-1} \sum_{k = 0}^N (\log
t)^{-k}\int_0^\infty e^{- x} \Big(\log x + i \pi - 2 \kappa
\Big)^k d x\ \ + \\  t^{-1} (\log t)^{-1} \cdot (\log t)^{-N - 1}
\int_0^\infty e^{- x} \frac{\Big(\log x + i \pi - 2 \kappa
\Big)^{N+1}}{1 - \frac{\log x + i \pi - 2 \kappa}{\log t} } d x .
\end{multline*}
The last integral here is bounded in $t$ as $t \to 0$. Since $N$
is arbitrary, it follows that
\[
f(t) \sim t^{-1} \sum_{k = 0}^\infty (\log t)^{-k-1}\int_0^\infty
e^{- x} \Big(\log x + i \pi - 2 \kappa \Big)^k d x.
\]
Therefore, since $\ell'(t) = f(t) + g(t)$, as $t \to 0$ we have
\[
\ell'(t) \sim t^{-1} \sum_{k = 0}^\infty (\log
t)^{-k-1}\int_0^\infty e^{- x} \Big(\log x + i \pi - 2 \kappa
\Big)^k d x + \sum_{k = 0}^\infty \gamma_k t^k.
\]
Integrating both sides, using that
\[
\int t^{-1} (\log t)^{-1} \, d t = \log | \log t|\quad , \quad
\int t^{-1} (\log t)^{-k - 1} d t = - \frac{1}{k} (\log t)^{-k} \
\ \text{for} \ \ k > 0,
\]
we get
\[
\ell(t) \sim \log |\log t| - \sum_{k = 1}^\infty \frac{1}{k} (\log
t)^{-k}\int_0^\infty e^{- x} \Big(\log x + i \pi - 2 \kappa
\Big)^k d x\\ + \sum_{k = 0}^\infty \delta_k t^k.
\]
Finally, in view of \eqref{int}, we see that as $t \to 0$,
\begin{multline*}
\frac{i}{2 \pi} \int_\gamma e^{-t \la} \frac{1}{(-\la)(\log (-\la)
- 2 \kappa)} d \la \sim \frac{1}{\pi} \Im \, \ell(t) \sim \\ -
\sum_{k = 1}^\infty \frac{1}{k \pi} (\log t)^{-k} \, \Im \left(
\int_0^\infty e^{- x} \Big(\log x + i \pi - 2 \kappa \Big)^k d x
\right) + \sum_{k = 0}^\infty \epsilon_k t^k .
\end{multline*}
\end{proof}

%%%%%%%%%%%%%%%%%%%%%%%%%%%%%%%%%%%%%%%%%%%%%%%%%%%%%%%%%%%%%%%%%
\section{The zeta determinant} \label{sec-det}
%%%%%%%%%%%%%%%%%%%%%%%%%%%%%%%%%%%%%%%%%%%%%%%%%%%%%%%%%%%%%%%%%

By Proposition \ref{prop-zeta}, we have
\[
\zeta(s,\Delta_\theta) = - \frac{e^{- 2 s \kappa} \sin \pi s}{\pi}
\log s + \zeta_\theta(s)
\]
where $\zeta_\theta(s)$ extends from $\Re s > 1/2$ to a
holomorphic function on $\C$ with poles at $s = 1/2 - k$ for $k =
0,1,2,\ldots$.  Since $\frac{e^{- 2 s \kappa} \sin \pi s}{\pi} = s
+ \mathcal{O}(s^2)$, it follows that
\[
\zeta_{\mathrm{reg}}(s,\Delta_\theta) := \zeta(s , \Delta_\theta)
+ s \log s
\]
has a derivative at $s = 0$.  Therefore, we can define
\[
{\det}_{\mathrm{reg}}
(\Delta_\theta):=\exp(-\zeta_{\mathrm{reg}}'(0,\Delta_\theta)),
\]
which is computed in this section.

Recall that $0 \leq \theta < \pi$ with $\theta \ne \frac\pi 2$.
The idea here is to make the first term in
\[
\zeta(s,\Delta_\theta) = \frac{\sin \pi s}{\pi} \int_{|t|}^\infty
x^{-2 s} \frac{d}{d x} \log F(i x)\, d x + \frac{1}{2 \pi i}
\int_{\gamma_t} \mu^{-2 s} \frac{F'(\mu)}{F(\mu)}\, d \mu
\]
regular at $s = 0$, as the second term (being entire) is already
regular at $s = 0$. In order to analytically continue the first
term, we add and subtract off the leading asymptotics of $F(ix)$.
Thus, recalling Lemma \ref{lem-asympF} (see \eqref{asymF0})
\[
F(i x) \sim C_0 (\log x - \kappa) x^{-\frac12} e^{xR} \left(
1+{\mathcal O} \left( \frac 1 x \right) \right) \quad \mbox{as
}x\to\infty, \] where $C_0 = - \frac{1}{\sqrt{2\pi R}}$, we
consider
\begin{align*}
\int_{|t|}^\infty  x^{-2 s} \frac{d}{d x} \log F(i x)\, d x   = &
\int_{|t|}^\infty x^{-2 s} \frac{d}{d x} \log \Big( \frac{F(i
x)}{C_0 (\log x - \kappa) x^{-\frac 12} e^{xR}} \Big)\, d x \\ & +
\int_{|t|}^\infty x^{-2 s} \frac{d}{d x} \log \Big( C_0 (\log x -
\kappa) x^{-\frac12} e^{xR} \Big)\, d x.
\end{align*}
The second integral can be computed explicitly:
\begin{align*}
\int_{|t|}^\infty x^{-2 s} \frac{d}{d x} & \log \Big( C_0 (\log x
-
\kappa) x^{-\frac12} e^{xR} \Big)\, d x\\
& = \int_{|t|}^\infty x^{-2 s} \frac{1}{x(\log x - \kappa)} \, d x
- \frac{|t|^{-2 s}}{4 s} + \frac{|t|^{- 2 s + 1}}{2 s - 1} R.
\end{align*}
Therefore,
\begin{multline*}
\zeta(s,\Delta_\theta ) = \frac{\sin \pi s}{\pi} \int_{|t|}^\infty
x^{-2 s} \frac{d}{d x} \log \Big( \frac{F(i x)}{C_0 (\log x -
\kappa) x^{-\frac12} e^{xR}} \Big)\, dx\\ + \frac{\sin \pi s}{\pi}
\int_{|t|}^\infty x^{-2 s} \frac{1}{x(\log x - \kappa)} \, d x  -
\frac{\sin \pi s}{\pi} \frac{|t|^{- 2 s}}{4 s} \\ + \frac{\sin \pi
s}{\pi} \frac{|t|^{- 2 s + 1}}{2 s - 1} R + \frac{1}{2 \pi i}
\int_{\gamma_t} \mu^{-2 s} \frac{F'(\mu)}{F(\mu)}\, d \mu .
\end{multline*}
Hence, as $\zeta_{\mathrm{reg}}(s,\Delta_\theta ) = \zeta(s ,
\Delta_\theta ) + s \log s$, we see that
\begin{multline*}
\zeta_{\mathrm{reg}}(s,\Delta_\theta ) = \frac{\sin \pi s}{\pi}
\int_{|t|}^\infty x^{-2 s} \frac{d}{d x} \log \Big( \frac{F(i
x)}{C_0 (\log x - \kappa) x^{-\frac12} e^{xR}} \Big)\, dx\\ +
\frac{\sin \pi s}{\pi} \int_{|t|}^\infty x^{-2 s} \frac{1}{x(\log
x - \kappa)} \, d x + s \log s
\\ - \frac{\sin \pi s}{\pi} \frac{|t|^{- 2 s}}{4
s} + \frac{\sin \pi s}{\pi} \frac{|t|^{- 2 s + 1}}{2 s - 1} R+
\frac{1}{2 \pi i} \int_{\gamma_t} \mu^{-2 s}
\frac{F'(\mu)}{F(\mu)}\, d \mu .
\end{multline*}
By Lemma \ref{lem-logint}, we have
\begin{align*}
\frac{\sin \pi s}{\pi} \int_{|t|}^\infty & x^{-2 s}
\frac{1}{x(\log x - \kappa)}\, d x \\ & = - \frac{e^{- 2 s \kappa}
\sin \pi s}{\pi} \log s - \frac{e^{- 2 s \kappa} \sin \pi s}{\pi}
\Big( \gamma + \log (2 (\log |t| - \kappa)) + \mathcal{O}(s)\Big)\\
& = - s \log s - s \Big( \gamma + \log (2 (\log |t| - \kappa)) +
\mathcal{O}(s \log s)\Big).
\end{align*}
Thus,
\begin{multline*}
\zeta_{\mathrm{reg}}(s,\Delta_\theta ) = \frac{\sin \pi s}{\pi}
\int_{|t|}^\infty x^{-2 s} \frac{d}{d x} \log \Big( \frac{F(i
x)}{C_0 (\log x - \kappa) x^{-\frac12} e^{xR}} \Big)\, dx\\ - s
\Big( \gamma + \log (2 (\log |t| - \kappa)) + \mathcal{O}(s^2 \log
s)\Big)
\\ - \frac{\sin \pi s}{\pi} \frac{|t|^{- 2 s}}{4
s} + \frac{\sin \pi s}{\pi} \frac{|t|^{- 2 s + 1}}{2 s - 1} R+
\frac{1}{2 \pi i} \int_{\gamma_t} \mu^{-2 s}
\frac{F'(\mu)}{F(\mu)}\, d \mu .
\end{multline*}
The first integral on the right is regular at $s = 0$ due to the
asymptotics found in Lemma \ref{lem-asympF}. Therefore, using that
\[
\frac{\sin (\pi s)}{\pi} \Big|_{s = 0} = 0 \ , \ \frac{d}{d s}
\frac{\sin (\pi s)}{\pi} \Big|_{s = 0} = 1 \ , \ \frac{\sin (\pi
s)}{\pi s} \Big|_{s = 0} = 1 \ , \ \frac{d}{d s} \frac{\sin (\pi
s)}{\pi s} \Big|_{s = 0} = 0,
\]
we see that
\begin{align*}
\zeta'_{\mathrm{reg}}(0,\Delta_\theta ) & = \int_{|t|}^\infty
\frac{d}{d x} \log \Big( \frac{F(i x)}{C_0 (\log x - \kappa)
x^{-\frac12} e^{xR}} \Big)\, dx\\ & - \Big( \gamma + \log (2 (\log
|t| - \kappa) \Big) + \frac12 \log |t| - |t|R - \frac{1}{\pi i}
\int_{\gamma_t} \log
\mu \frac{F'(\mu)}{F(\mu)}\, d \mu \\
& = - \log \Big( \frac{F(i |t|)}{C_0} \Big)  - \gamma - \log 2 -
\frac{1}{\pi i} \int_{\gamma_t} \log \mu \frac{F'(\mu)}{F(\mu)}\,
d \mu .
\end{align*}
Therefore,
\begin{equation} \label{detreg}
{\det}_{\mathrm{reg}} (\Delta_\theta ) = 2 e^\gamma
\frac{F(t)}{C_0} \cdot \exp\Big( \frac{1}{\pi i} \int_{\gamma_t}
\log \mu \frac{F'(\mu)}{F(\mu)}\, d \mu \Big).
\end{equation}
This formula is derived, a priori, when $t$ is on the upper half
of the imaginary axis. However, the right side is a
\emph{holomorphic} function of $t \in \dD$, where $\dD$ is the set
of complex numbers minus the negative real axis and the zeros of
$F(\mu)$. Therefore \eqref{detreg} holds for all $t \in \dD$.
Here, we recall that $\gamma_t$ is any curve in $\dD$ from $t$ to
$-t$. As before, the trick now is to let $t \to 0$ in
\eqref{detreg}.

\begin{figure}
\centering \includegraphics{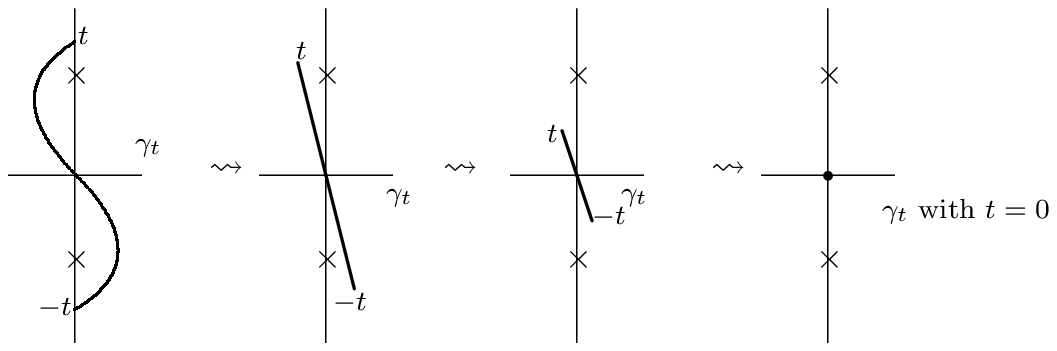}
\caption{The contour $\gamma_t$ as we let $t \to 0$ in $\dD$ from
the upper half plane.} \label{fig-contstrangedeform}
\end{figure}

First, assume that $\log R - \tan\theta \neq 0$ so that
$\Delta_\theta$ has no zero eigenvalue by Theorem
\ref{thm-eigenv}. We determine the limit as $t \to 0$ of the
exponential $\exp\big( \frac{1}{\pi i} \int_{\gamma_t} \log \mu
\frac{F'(\mu)}{F(\mu)}\, d \mu \big)$. Let's take $t \to 0$ in
$\dD$ from the upper half plane as shown in Figure
\ref{fig-contstrangedeform}.
In view of Figure \ref{fig-contstrangedeform}, it follows that
\[
\exp\Big(  \frac{1}{\pi i} \int_{\gamma_t} \log \mu
\frac{F'(\mu)}{F(\mu)}\, d \mu \Big) \to \exp\Big( 0 \Big) = 1.
\]
Recalling \eqref{Fasymp00}, as $\mu \to 0$, we have
\[
F(\mu) \, \sim\,  (\log R- \tan \theta )+ \frac14 \mu^2 R^2( 1 +
\tan \theta - \log R) + \mathcal{O}(\mu^4) .
\]
In this case $F(0) = \log R- \tan \theta$. In conclusion, taking
$t \to 0$ in \eqref{detreg}, we see that
\begin{equation}\label{e:det2}
{\det}_{\mathrm{reg}} (\Delta_\theta ) = 2 \sqrt{2 \pi R} e^\gamma
(\tan \theta -\log R).
\end{equation}

Second,  assume now that $\log R - \tan \theta = 0$ so that as
$\mu \to 0$, we have
\[
F(\mu)\, \sim\,  \frac14 \mu^2 R^2 ( 1 + \mathcal{O}(\mu^2)) .
\]
Let us put
\[
\widetilde{F}(\mu) := \frac{F(\mu)}{\mu^2};
\]
then $\widetilde{F}(\mu)$ is nonzero at $\mu = 0$ with value
$\frac{R^2}4$, and
\[
\zeta(s,\Delta_\theta ) = \frac{\sin \pi s}{\pi} \int_{|t|}^\infty
x^{-2 s} \frac{d}{d x} \log \widetilde{F}(i x)\, d x + \frac{1}{2
\pi i} \int_{\gamma_t} \mu^{-2 s}
\frac{\widetilde{F}'(\mu)}{\widetilde{F}(\mu)}\, d \mu.
\]
By Lemma \ref{lem-asympF} (see \eqref{asymF0}), we have
\[
\widetilde{F}(i x) \sim \frac{C_0 (\log x - \kappa) x^{-\frac12}
e^{xR}}{-x^2} = \widetilde{C}_0 (\log x - \kappa) x^{-\frac 52 }
e^{xR} , \quad \text{where}\ \ \widetilde{C}_0 =
\frac{1}{\sqrt{2\pi R }} .
\]
Now following the argument above used to prove \eqref{detreg}, we
can show
\[
{\det}_{\mathrm{reg}} (\Delta_\theta ) = 2 e^\gamma
\frac{\widetilde{F}(t)}{\widetilde{C}_0} \cdot \exp\Big(
\frac{1}{\pi i} \int_{\gamma_t} \log \mu
\frac{\widetilde{F}'(\mu)}{\widetilde{F}(\mu)}\, d \mu \Big).
\]
Finally, taking $t \to 0$ as we did before, yields in the
$\tan\theta = {\log R}$ case, the result
\[
{\det}_{\mathrm{reg}} (\Delta_\theta ) = 2 e^\gamma
\frac{R^2}{4\widetilde{C}_0}  = \frac {R^2} 2 e^\gamma \sqrt{2 \pi
R } = \sqrt{\frac{\pi R}{2}} e^\gamma R^2.
\]

%%%%%%%%%%%%%%%%%%%%%%%%%%%%%%%%%%%%%%%%%%%%%%%%%%%%%%%%%%%%%%%%%
\section{General boundary conditions at $r = R$}
\label{sec-bdycond}
%%%%%%%%%%%%%%%%%%%%%%%%%%%%%%%%%%%%%%%%%%%%%%%%%%%%%%%%%%%%%%%%%

We now prove Theorem \ref{thm-main}. Let's briefly recall the
set-up. Let $0 \leq \theta_1, \theta_2 < \pi$ with $\theta_1 \ne
\pi/2$ and put $L := L_{\theta_1} \oplus L_{\theta_2}$. Then as a
consequence of Theorem \ref{thm-sae}, we know that
\[
\Delta_L := \Delta : \dD_L \to L^2([0,R])
\]
is self-adjoint, where
\[
\dD_L = \{ \phi \in \dD_\max(\Delta)\, |\, \cos \theta_1 \,
c_1(\phi) + \sin \theta_1\, c_2(\phi) = 0\ , \ \cos \theta_2 \,
\phi '(R) + \sin \theta_2\, \phi (R) = 0 \} .
\]
The trick to proving Theorem \ref{thm-main} is to write the
resolvent $(\Delta_L - \la)^{-1}$ in terms of $(\Delta_{\theta_1}
- \la)^{-1}$ (same self-adjoint condition at $r = 0$ but with the
Dirichlet condition at $r = R$). To do so, let $\varrho(r) \in
C^\infty((-\infty ,\infty))$ be a non-decreasing function such
that $\varrho(r) = 0$ for $r \leq 1/4$ and $\varrho(r) = 1$ for $r
\geq 3/4$. Given any real numbers $\alpha < \beta$, we define
\begin{equation} \label{rhoab}
\varrho_{\alpha,\beta}(r) := \varrho((r-\alpha)/(\beta-\alpha)).
\end{equation}
Then $\varrho_{\alpha,\beta}(r) = 0$ on a neighborhood of $\{r
\leq \alpha\}$ and $\varrho_{\alpha,\beta}(r) = 1$ on a
neighborhood of $\{r \geq \beta\}$. We define
\begin{equation}
\begin{array}{l}
\psi_1(r)= \varrho_{R/2,3R/4}(r),\quad \psi_2(r)= 1-\psi_1(r),\\
\varphi_1(r)= \varrho_{R/4,R/2}(r),\quad \varphi_2(r)= 1 -
\varrho_{3R/4,R}(r).\end{array}\label{psiphi}
\end{equation}

Let $\Delta' := - \frac{d^2}{d r^2} - \frac{1}{4 r^2}$ over
$[\frac R 4 , R]$ with the Dirichlet condition at $r = R/4$ and
the condition $\cos \theta_2 \, \phi '(R) + \sin \theta_2\, \phi
(R) = 0$ at $r = R$; note that since $r \geq R/4$ over $[\frac R 4
, R]$, the operator $\Delta'$ is a true smooth elliptic operator
over this interval with no singularities. We define
\begin{equation} Q(\la) := \varphi_1 (\Delta' - \la)^{-1} \psi_1 + \varphi_2
(\Delta_{\theta_1} - \la)^{-1} \psi_2.\label{Q}
\end{equation}
It follows that $Q(\la)$ maps into the domain $\dD_L$ of
$\Delta_L$, and
\begin{align*}
(\Delta_L - \la) Q(\la) & = (\Delta_L - \la) \varphi_1 (\Delta' -
\la)^{-1} \psi_1 + (\Delta_L - \la) \varphi_2 (\Delta_{\theta_1} -
\la)^{-1} \psi_2 \\
& = \varphi_1 (\Delta' - \la) (\Delta' - \la)^{-1} \psi_1 +
\varphi_2 (\Delta_{\theta_1} - \la)
(\Delta_{\theta_1} - \la)^{-1} \psi_2 + K_0(\la) \\
& = \psi_1 + \psi_2 + K_0(\la) = \Id + K_0(\la) ,
\end{align*}
where
\[
K_0 (\la) = [\Delta ,\varphi_1] (\Delta' - \la)^{-1} \psi_1 +
[\Delta ,\varphi_2] (\Delta_{\theta_1} - \la)^{-1} \psi_2,
\]
Because the supports of $[\Delta ,\varphi_i]$ and $\psi_i$, where
$i = 1,2$, are disjoint, using the explicit formula \eqref{res}
for the resolvent $(\Delta_{\theta_1} - \la)^{-1}$ and the
properties of the resolvent of $(\Delta' - \la)^{-1}$ found in the
work of Seeley \cite{SeRAE69} it is straightforward to check that
$K_0(\la)$ is trace-class operator that vanishes to infinite order
as $|\la| \to \infty$ for $\la$ in any sector $\Lambda$ of $\C$
not intersecting the positive real axis; we shall fix such a
sector $\Lambda$ from now on. In particular, forming the Neumann
series, it follows that $\Id + K_0(\la)$ is invertible for $|\la|$
large with $\la \in \Lambda$ with
\[
(\Id + K_0(\la))^{-1} = \Id + K (\la),
\]
where $K(\la)$ has the same properties as $K_0(\la)$. Thus,
multiplying both sides of $(\Delta_L - \la) Q(\la) = \Id +
K_0(\la)$ by $\Id + K(\la)$, we see that
\[
(\Delta_L - \la)^{-1} = Q(\la) + Q(\la) K(\la).
\]
Therefore, as $|\la| \to \infty$ for $\la \in \Lambda$, we see
that Proposition \ref{prop-res} holds also for $\Tr (\Delta_L -
\la)^{-1}$. Now using the resolvent asymptotics, we can proceed to
copy the proofs of Proposition \ref{prop-hk} and \ref{prop-zeta}.
The proof of Theorem \ref{thm-main} is now complete.

%%%%%%%%%%%%%%%%%%%%%%%%%%%%%%%%%%%%%%%%%%%%%%%%%%%%%%%%%%%%%%%%%
\appendix
%%%%%%%%%%%%%%%%%%%%%%%%%%%%%%%%%%%%%%%%%%%%%%%%%%%%%%%%%%%%%%%%%

%%%%%%%%%%%%%%%%%%%%%%%%%%%%%%%%%%%%%%%%%%%%%%%%%%%%%%%%%%%%%%%%%
\section{The resolvent with Dirichlet conditions at $r =
R$} \label{sec-appendix}
%%%%%%%%%%%%%%%%%%%%%%%%%%%%%%%%%%%%%%%%%%%%%%%%%%%%%%%%%%%%%%%%%

In this Appendix, we compute the trace of the resolvent by
explicitly finding the Schwartz kernel of the Bessel function. To
do so, recall that the resolvent kernel of the differential
operator $\Delta_\theta - \mu^2$ with given boundary conditions at
$r=0$ and $r = R$ can be expressed as follows (see Lemma 4.1 in
\cite{BrJ-SeR88} or \cite[Sec.\ 3.3]{BHiF92} for an elementary
account):
\[
\frac{-1}{W(p,q)} \begin{cases} p(r,\mu) \, q(s,\mu) & \text{for}
\quad
r\leq s \\
p(s,\mu) \, q(r,\mu) & \text{for} \quad r\geq s,
\end{cases}
\]
where $p(r,\mu)$ and $q(r,\mu)$ are solutions of $\big( \Delta_L -
\mu^2 \big) \phi = 0$ satisfying the given boundary conditions at
$r=0$ and $r = R$, respectively, and where $W(p,q)$ is the
Wronskian of $(p,q)$. Recall that the general solution to $\big(
\Delta_L - \mu^2 \big) \phi = 0$ is
\[
\phi = c_1(\phi)\, r^{1/2} \, J_0(\mu r) +c_2(\phi)\, r^{1/2} \,
\Big(\frac{\pi}{2}  Y_0(\mu r) - (\log \mu - \log 2 + \gamma)\,
J_0(\mu r)\Big),
\]
To satisfy the boundary condition at $r = 0$, we must have (see
\eqref{bdycond0}):
\[
\cos \theta \, c_1(\phi) + \sin \theta \, c_2 (\phi) = 0.
\]
Thus, we can take $c_2(\phi) = 1$ and $c_1(\phi) = - \tan \theta$;
this gives
\begin{equation} \label{pformula}
p(r,\mu) = r^{1/2} \, \Big(\frac{\pi}{2}  Y_0(\mu r) - (\log \mu -
\kappa)\, J_0(\mu r)\Big),
\end{equation}
where $\kappa = \log 2 - \gamma - \tan \theta$. To determine
$q(r,\mu)$ we use the less fancy formulation of the general
solution:
\[
\phi = C_1 \, r^{1/2} J_0(\mu r) + C_2 \, r^{1/2} Y_0(\mu r).
\]
To satisfy the Dirichlet condition at $r = R$, we therefore take
\begin{equation} \label{qformula}
q(r,\mu) = r^{1/2}\, \Big( Y_0(\mu R)\, J_0(\mu r) - J_0(\mu R)\,
Y_0(\mu r) \Big).
\end{equation}
The Wronskian is easily computed using that $W (r^{1/2} J_0(\mu r)
, r^{1/2} Y_0(\mu r)) = \frac{2}{\pi}$ (see \cite{UWolfBesselJ}):
\begin{align*}
W(p,q) = \frac{\pi}{2}\, Y_0(\mu R) \, W ( r^{1/2} Y_0(\mu r) , &
r^{1/2} J_0(\mu r))\\ & + (\log \mu - \kappa) J_0(\mu R) \, W
(r^{1/2} J_0(\mu r) , r^{1/2} Y_0(\mu r))\\
& = - Y_0(\mu R) + \frac{2}{\pi} (\log \mu - \kappa) J_0(\mu R).
\end{align*}
Therefore,
\begin{equation} \label{res}
(\Delta_\theta - \mu^2)^{-1} (r,s) = \frac{1}{F(\mu)}
\begin{cases} p(r,\mu) \, q(s,\mu) & \text{for}
\quad r \leq s \\
p(s,\mu) \, q(r,\mu) & \text{for} \quad r\geq s,
\end{cases}
\end{equation}
where $p$ and $q$ are given in \eqref{pformula} and
\eqref{qformula}, respectively, and where
\[
F(\mu) : = Y_0(\mu R) - \frac{2}{\pi} (\log \mu - \kappa) J_0(\mu
R ).
\]
We now need to compute $\int_0^R p(r,\mu)\, q(r,\mu)\, dr$; that
is,
\begin{equation} \label{tracecalc}
\begin{split}
\int_0^R r \, \Big(\frac{\pi}{2} &  Y_0(\mu r) - (\log \mu -
\kappa)\, J_0(\mu r)\Big) \, \Big( Y_0(\mu R)\, J_0(\mu r) -
J_0(\mu R)\, Y_0(\mu r) \Big)\, dr  \\
& = \frac{\pi}{2} Y_0(\mu R) \int_0^R r \, Y_0(\mu r) \, J_0(\mu
r)
\, dr - \frac{\pi}{2} J_0(\mu R) \int_0^R r \, Y_0(\mu r)^2 \, dr \\
& + (\log \mu - \kappa) J_0(\mu R) \int_0^R r J_0(\mu r) \,
Y_0(\mu
r) \, dr \\
& - (\log \mu - \kappa) Y_0(\mu R) \int_0^R r \, J_0(\mu r)^2 \,
dr.
\end{split}
\end{equation}
We next use the indefinite integrals
\begin{align*}
\int r J_0(\mu r)^2\, d r & = \frac{r^2}{2} \big( J_0 (\mu r)^2 +
J_1(\mu r)^2 \big)\\
\int r Y_0(\mu r)^2\, d r & = \frac{r^2}{2} \big( Y_0 (\mu r)^2 +
Y_1(\mu r)^2 \big)\\
\int r Y_0(\mu r)\, J_0(\mu r)\, d r & = \frac{r^2}{2} \big( Y_0
(\mu r) J_0 (\mu r) + Y_1 (\mu r) J_1(\mu r) \big) ,
\end{align*}
which we need to evaluate between $r=0$ and $r=R$. Recalling
\eqref{Jvdef}, $z J_0(z)$ and $z J_1(z)$ vanish at $z = 0$. Also,
by \eqref{Y0def} $z Y_0(z)$ also vanishes at $z = 0$. However, it
is a remarkable fact, which may be easily overlooked, that since
\[
\frac{\pi}{2} Y_1(z) = - \frac{1}{z} J_0(z) + \big( \log z - \log
2 + \gamma \big) J_1(z) - \frac12 z \sum_{k = 1}^\infty \frac{k
H_k (- \frac14 z^2 )^{k-1}}{(k!)^2} ,
\]
where we used that $Y_1(z) = - Y_0'(z)$ and $J_1(z) = - J_0'(z)$
from \cite[p.\ 361]{BAbM-StI92}, we have
\[
\Big( z Y_1(z) \Big)\Big|_{z = 0} = - \frac{2}{\pi} \qquad
\Longrightarrow \qquad \Big( z^2 Y_1(z)^2 \Big)\Big|_{z = 0} =
\frac{4}{\pi^2} .
\]
 Therefore,
\begin{align*}
\int_0^R r J_0(\mu r)^2\, d r & = \frac{R^2}{2} \big( J_0 (\mu
R)^2 +
J_1(\mu R)^2 \big)\\
\int_0^R r Y_0(\mu r)^2\, d r & = \frac{R^2}{2} \Big( Y_0 (\mu
R)^2
+ Y_1(\mu R)^2 \Big) - \frac 1 2 \Big( r^2 Y_1(\mu r)^2 \Big) \Big|_{r = 0} \Big)\\
& = \frac{R^2}{2} \big( Y_0 (\mu R)^2 +
Y_1(\mu R)^2 \big) - \frac{2}{\pi^2 \mu^2} \\
\int_0^R r Y_0(\mu r)\, J_0(\mu r)\, d r & = \frac{R^2}{2} \big(
Y_0 (\mu R) J_0 (\mu R) + Y_1 (\mu R) J_1(\mu R) \big) .
\end{align*}
Plugging these integrals into \eqref{tracecalc} and using the
identity \cite[p.\ 360]{BAbM-StI92}
\[
J_1(z) \, Y_0(z) - J_0(z)\, Y_1(z) = \frac{2}{\pi z}
\]
to simplify the expression obtained, we eventually arrive that
\begin{align*}
\int_0^R p(r,\mu)\, q(r,\mu)\, dr & = \frac{R}{2 \mu} \Big(
Y_1(\mu R) - \frac{2}{\pi}(\log \mu - \kappa) J_1(\mu R) \Big) +
\frac{1}{\pi \mu^2} J_0(\mu R)\\
& = \frac{R}{2 \mu} \Big( Y_1(\mu R) - \frac{2}{\pi}(\log \mu -
\kappa) J_1(\mu R) + \frac{2}{\pi \mu R} J_0(\mu R) \Big).
\end{align*}
Using the fact that $J_0'(z) = - J_1(z)$ and $Y_0'(z) = - Y_1(z)$,
we can write this as
\[
\int_0^R p(r,\mu)\, q(r,\mu)\, dr = -  \frac{1}{2 \mu} \frac{d}{d
\mu} \Big( Y_0 (\mu R) - \frac{2}{\pi}(\log \mu - \kappa) J_0(\mu
R ) \Big) = -  \frac{1}{2 \mu} \frac{d}{d \mu} F(\mu) .
\]
where we recall that
\[
F(\mu) : = Y_0(\mu R) - \frac{2}{\pi} (\log \mu - \kappa) J_0(\mu
R ).
\]
Since (see \eqref{res})
\[
(\Delta_\theta - \mu^2)^{-1} (r,s) = \frac{1}{F(\mu)}
\begin{cases} p(r,\mu) \, q(s,\mu) & \text{for}
\quad r \leq s \\
p(s,\mu) \, q(r,\mu) & \text{for} \quad r\geq s,
\end{cases}
\]
we have proved the following theorem,
%which is in agreement with
%Theorem \ref{thm-res}.
\begin{theorem} \label{thm-traceres}
With $F(\mu) : = Y_0(\mu R) - \frac{2}{\pi} (\log \mu - \kappa)
J_0(\mu R)$, we have
\begin{align*}
\Tr( \Delta_\theta - \mu^2)^{-1} & = -  \frac{1}{2 \mu}
\frac{1}{F(\mu)} \frac{d}{d \mu} F(\mu)\\ & = - \frac{1}{2 \mu}
\frac{d}{d \mu} \log F(\mu).
\end{align*}
\end{theorem}
This theorem has been used to analyze the zeta function,
resolvent, and heat kernel of $ \Delta_\theta$ in Sections
\ref{sec-zeta}--\ref{sec-heat}.

\section*{Acknowledgements}
KK was supported in part by funds from the Baylor University
Research Committee and by the Max-Planck-Institute for Mathematics
in the Sciences (Leipzig, Germany).

%----------------------------------------------------------------
\bibliographystyle{amsplain}
\def\cprime{$'$} \def\polhk#1{\setbox0=\hbox{#1}{\ooalign{\hidewidth
  \lower1.5ex\hbox{`}\hidewidth\crcr\unhbox0}}}
\providecommand{\bysame}{\leavevmode\hbox
to3em{\hrulefill}\thinspace}
\providecommand{\MR}{\relax\ifhmode\unskip\space\fi MR }
\providecommand{\MRhref}[2]{%
  \href{http://www.ams.org/mathscinet-getitem?mr=#1}{#2}
} \providecommand{\href}[2]{#2}

%----------------------------------------------------------------

\end{document}